\documentclass[aps,floatfix,superscriptaddress,twocolumn,tightenlines,amsmath,amssymb,nofootinbib,10pt,longbibliography,prl]{revtex4-2}

\usepackage[utf8]{inputenc}
\usepackage[T1]{fontenc}
\usepackage[mathscr]{euscript}
\usepackage{graphicx}
\usepackage[dvipsnames]{xcolor}
\usepackage{amsthm}
\usepackage{psfrag}
\usepackage{ifsym}
\usepackage{dsfont}
\usepackage{enumerate}
\usepackage{amsfonts}
\usepackage{multirow}
\usepackage{mathtools}
\usepackage{amsmath}
\usepackage{hyperref}
\hypersetup{colorlinks=true, linkcolor=red, citecolor=blue}
\usepackage[sort&compress]{natbib}
\usepackage[separate-uncertainty=true]{siunitx}
\usepackage{comment}
\usepackage[normalem]{ulem}
\usepackage[sc,osf]{mathpazo}
\usepackage{bm}
\usepackage{bbm}
\usepackage{xr}

\def\1{\mathbf{1}}
\def\0{\mathbf{0}}

\begin{document}

\title{Multiclass Portfolio Optimization via Variational Quantum Eigensolver with Dicke State Ansatz}

\author{J. V. S. Scursulim}
\affiliation{Instituto de Ciência e Tecnologia Itaú}

\author{Gabriel M. Langeloh}
\affiliation{Instituto de Ciência e Tecnologia Itaú}
\author{Victor L. Beltran}
\affiliation{Instituto de Ciência e Tecnologia Itaú}
\author{Samuraí Brito}
\affiliation{Instituto de Ciência e Tecnologia Itaú}


\date{\today}

\begin{abstract}
Combinatorial optimization is a fundamental challenge in various domains, with portfolio optimization standing out as a key application in finance. Despite numerous quantum algorithmic approaches proposed for this problem, most overlook a critical feature of realistic portfolios: diversification. In this work, we introduce a novel quantum framework for multiclass portfolio optimization that explicitly incorporates diversification by leveraging multiple parametrized Dicke states, simultaneously initialized to encode the diversification constraints, as an ansatz of the Variational Quantum Eigensolver. A key strength of this ansatz is that it initializes the quantum system in a superposition of only feasible states, inherently satisfying the constraints. This significantly reduces the search space and eliminates the need for penalty terms. In addition, we also analyze the impact of different classical optimizers in this hybrid quantum-classical approach. Our findings demonstrate that, when combined with the CMA-ES optimizer, the Dicke state ansatz achieves superior performance in terms of convergence rate, approximation ratio, and measurement probability. These results underscore the potential of this method to solve practical, diversification-aware portfolio optimization problems relevant to the financial sector.

\end{abstract}

\maketitle

\section{Introduction}
\label{Sec: Introduction}

Quantum Computing (QC) represents a groundbreaking technology poised to solve problems that lie beyond the capabilities of classical computers. It operates by manipulating quantum systems and harnessing unique properties such as superposition, interference, and entanglement. This approach enables the execution of advanced algorithms designed to address complex challenges with remarkable efficiency. The potential of quantum computing spans multiple fields, promising significant impacts across industries and scientific disciplines; some of them include: quantum chemistry and material science \cite{mcardle2020quantum, bauer2020quantum, von2021quantum}, machine learning \cite{biamonte2017quantum, liu2018quantum, schuld2019quantum, cong2019quantum, cerezo2022challenges, jerbi2023quantum}, finance \cite{egger2020quantum, ramos2021quantum, mugel2022dynamic, buonaiuto2023best, herman2023quantum, wilkens2023quantum, naik2025portfolio, Thakkar2024}, and optimization \cite{jarret2018improved, campbell2019applying, montanaro2020quantum, egger2021warm, magann2022feedback, abbas2024challenges, finvzgar2024quantum}.

Optimization plays a fundamental role across various domains, with classic examples including the traveling salesman problem \cite{junger1995traveling}, vehicle routing \cite{toth2002vehicle}, bin packing \cite{martello2000three}, and portfolio optimization \cite{mansini2015linear}—the central focus of this article. These problems are traditionally tackled using classical techniques such as mixed-integer programming \cite{junger200950}, approximation algorithms and metaheuristics \cite{gonzalez2007handbook}, and neural networks \cite{smith1999neural, prates2019learning}. However, due to their computational complexity, these methods tend to have limitations in either their running time or solution quality as the number of variables increases.

Portfolio optimization exemplifies this challenge, especially in multiclass scenarios that require allocation across diverse asset categories (e.g., equities, bonds, commodities). While classical approaches like the Markowitz mean-variance model \cite{markowitz1952portfolio} perform well for small-scale instances, multiclass formulations introduce combinatorial and equality constraints that lead to mixed-integer quadratic programs (MIQPs), which are hard to solve optimally, particularly in real-time or resource-limited settings.

In response to these limitations, Variational Quantum Algorithms (VQAs) \cite{cerezo2021variational, preskill2018quantum} have emerged as promising tools for near-term quantum advantage. Designed for Noisy Intermediate-Scale Quantum (NISQ) devices, VQAs, especially the Variational Quantum Eigensolver (VQE) \cite{vqe2014},  have demonstrated potential in solving combinatorial problems by mapping them to Ising Hamiltonians and approximating the ground state via parameterized quantum circuits. The effectiveness of this approach heavily depends on the choice of ansatz, particularly when handling complex constraints such as those in multiclass portfolio optimization.

To that end, this paper proposes the use of a Dicke state-based ansatz within the VQE framework to efficiently handle multiclass portfolio optimization, incorporating a realistic feature in this problem related to diversification of the investment. Our contributions are threefold:  
(i) \textit{Modeling advantage} — we formulate a multiclass portfolio optimization problem suitable for quantum encoding by introducing a parameterized Dicke-state ansatz, in which diversification constraints are inherently satisfied through state preparation, eliminating the need for penalty calibration;  (ii) \textit{Search-space advantage} — by employing the Dicke state within the VQE framework, we drastically reduce the effective search space, restricting the sampling and optimization to the feasible manifold, which improves sample efficiency and convergence stability; and  (iii) \textit{Empirical advantage} — in our simulations, the Dicke ansatz combined with CMA-ES achieved higher approximation ratios and more frequent identification of the global optimum than standard ansatzes at comparable parameter counts. We make no claim of asymptotic quantum speedup but demonstrate meaningful structural and practical improvements within the variational paradigm.

This paper is organized as follows: the first section introduces the problem of multiclass portfolio optimization. Subsequently, we provide an overview of VQE and the standard ansatz. We then describe the problem formulation, the Dicke state ansatz, and the methodology employed. Finally, we present the numerical results, empirical findings, and discussions. The paper concludes with a summary of the main contributions and outlines directions for future research in quantum finance.

\section{Multiclass portfolio optimization}
\label{Sec: Multiclass}
\begin{figure}[!htb]
\begin{center}
\includegraphics[scale=.42]{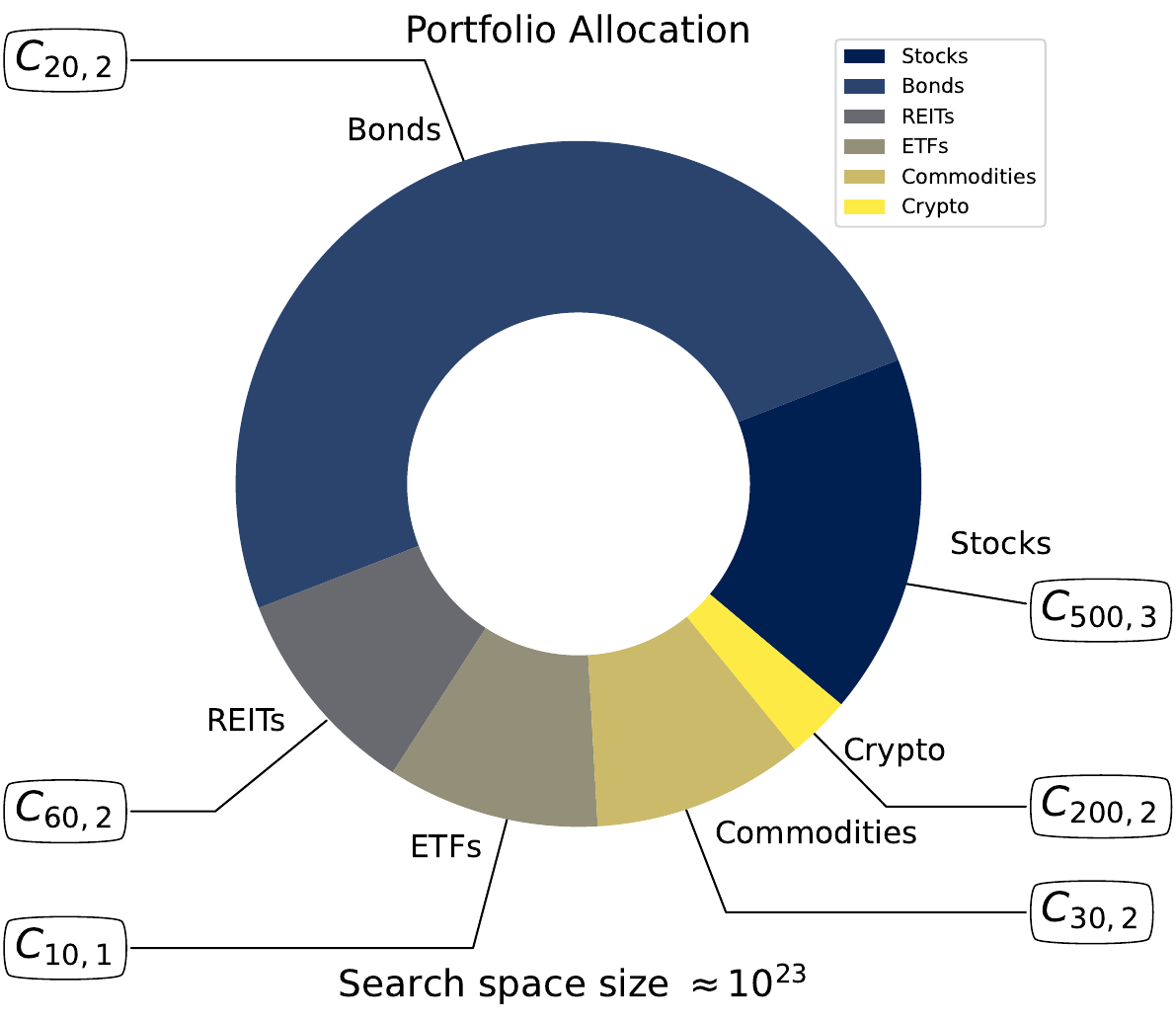}
\end{center}
\caption{Illustrative example of a multiclass portfolio optimization problem with a predefined asset class allocation to ensure diversification aiming the reduction of market risk. The total number of assets considered is $820$, distributed as follows: $500$ stocks, $200$ cryptocurrencies, $30$ commodities, $10$ ETFs, $60$ REITs and $20$ Bonds. For each class there are $C_{n,k}$ possible portfolios, where $n$ represents the total number of assets in the class and $k$ is the predefined number of assets to be selected. The goal of this portfolio optimization is to find the best set of assets that will produce a portfolio that satisfies the constraints maximizing the return and minimizing the risk.}

\label{fig1}
\end{figure}

In practical financial scenarios, optimizing solely for return and risk is insufficient. A realistic and robust portfolio must also incorporate diversification constraints. This involves not only selecting a larger number of assets, but also ensuring representation across different asset classes, such as stocks, bonds, and other financial instruments (see Fig.\ref{fig1}).

Diversification reduces exposure to specific market segments, enhances portfolio resilience, and aligns with best practices in risk management. In its classical form, the objective is to select a subset of assets that minimizes portfolio risk while maximizing expected return, under a given level of risk aversion \cite{markowitz1952portfolio}, as formulated below
\begin{align}
    \min_{x \ \in \ \{0,1\}^n} \ & qx^T\Sigma x - (1-q)x^T\mu + r_f, \label{mean_variance}\\
    s.t. \ & Ax=b \nonumber.
\end{align}
where $x$ is a binary decision vector such that an entry equal to 1 indicates the inclusion of a corresponding asset in the portfolio, $\Sigma$ is the covariance matrix of asset returns, capturing the portfolio’s overall risk, while $\mu$ represents the expected return vector. The parameter $r_f$ denotes the risk-free rate, such as the return of US Treasury bonds. The matrix $A_{m \times n}$ encodes the linear constraints that govern portfolio selection, where $m$ is the number of constraint equations and typically corresponds to the number of asset classes or diversification. In the example of Fig.~\ref{fig1}, $a_{ij} = 1$ whenever asset $j$ belongs to class $i$ and $a_{ij} = 0$ otherwise. To simplify and reduce the number of experiment parameters, we set $q=0.5$. 

This formulation enables the inclusion of constraints that promote diversification, by limiting or enforcing the number of assets selected within each predefined class. Such constraints are essential for constructing well-diversified portfolios, which are less exposed to specific sector or asset risks, and thus more resilient to market fluctuations.

\section{Variational Quantum Algorithms and Ansatz}
\label{Sec: VQE}

\begin{figure}[!htb]
\begin{center}
\includegraphics[scale=.25]{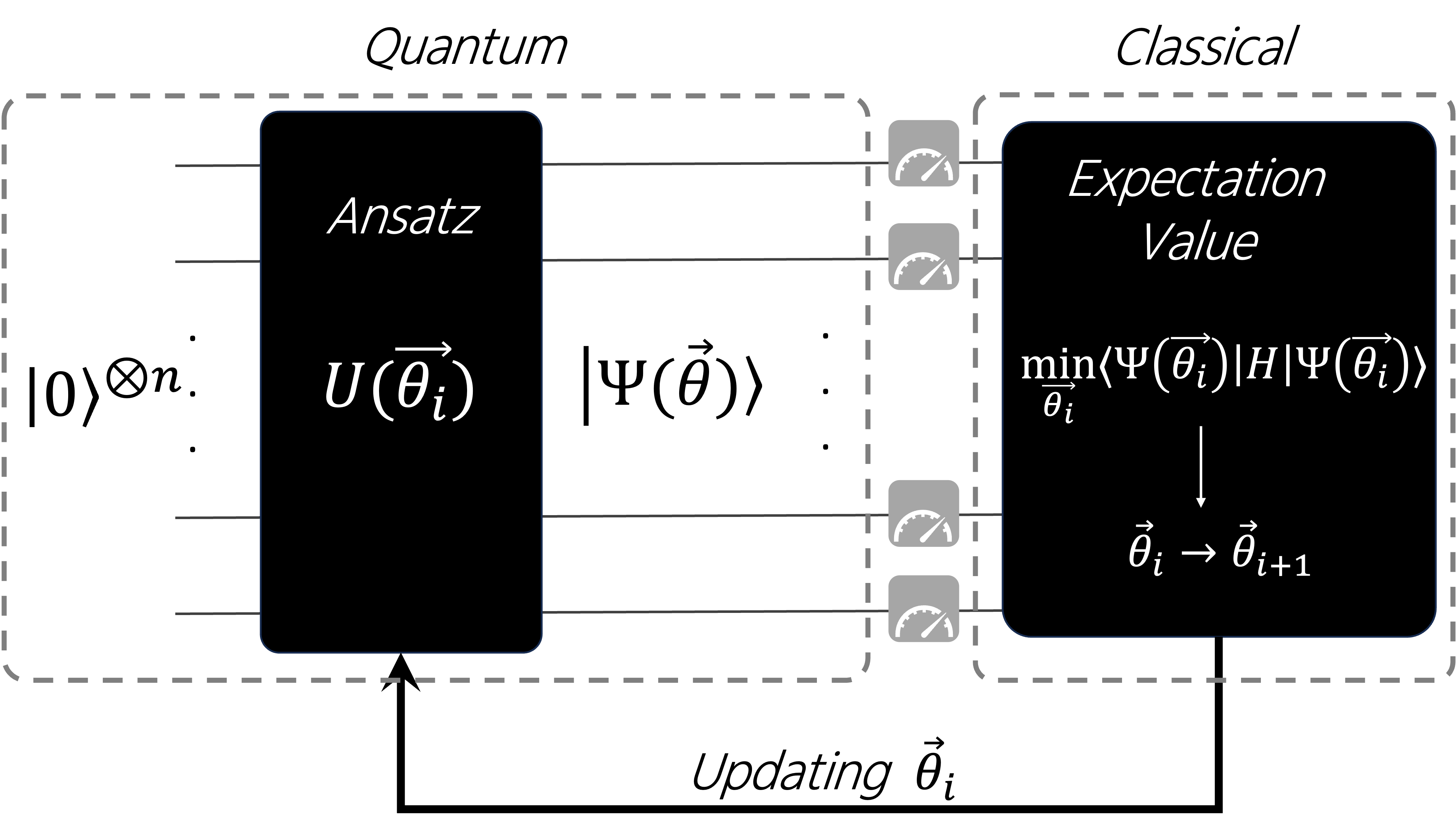}
\end{center}
\caption{An illustrative example of a VQE routine, which is defined by a quantum state preparation conducted on a quantum device or simulator, followed by an optimization process in a classical computer. In the quantum routine, we start with a quantum state where all qubits are in state $\vert0\rangle$, this initial state evolves according to the unitary $U(\vec{\theta_i})$, which defines ansatz structure and receives a set of parameters that will define the states probability distribution extracted from a certain number of measurements. The classical routine is focused on updating the set of parameters $\vec{\theta_i}$, in order to minimize the expectation value of the Hamiltonian that encodes the optimization problem. This process is repeated until the maximum number of iterations or when other stopping criteria are achieved.}
\label{fig2}
\end{figure}

The primary VQAs for combinatorial optimization are VQE \cite{vqe2014} and the Quantum Approximate Optimization Algorithm (QAOA)  \cite{farhi2014quantum}. In this work, we focus only on VQE in the context of multiclass portfolio optimization (see Fig.\ref{fig2}). Both approaches utilize a parameterized quantum circuit, commonly referred to as an ansatz, together with a classical optimization routine. Their objective is to minimize the expectation value of the problem Hamiltonian ($H$) which encodes the specific combinatorial optimization problem to be solved, as showed in the equation below:
\begin{eqnarray}
    \min_{\vec{\theta} \ \in \ \mathbb{R}^n} \langle\psi(\vec{\theta})\vert H \vert\psi(\vec{\theta})\rangle. \label{vqe_obj_fun}
\end{eqnarray}
The expectation value defined in equation \eqref{vqe_obj_fun} is lower bounded by the minimum eigenvalue of the Hamiltonian $H$, known as the energy of the ground state $E_0$. The primary goal of this approach is to identify an appropriate ansatz along with an optimal set of parameters $\vec{\theta^*}$, such that the expected value computed $\langle\psi(\vec{\theta^*})|H|\psi(\vec{\theta^*})\rangle$ is equal or close to $E_0$. Quantum mechanics guarantees the existence of this lower bound \cite{griffiths2019introduction}, although its exact value is generally unknown beforehand. Thus, the variational method provides a practical way to approximate both the ground-state wavefunction and its corresponding energy. 

The main difference between VQE and QAOA is that the latter is a special case of the former, once the QAOA ansatz has a well-defined structure based on the adiabatic theorem \cite{griffiths2019introduction}, which, in general, leads to deeper quantum circuits. There is a trade-off between these algorithms, VQE offers a shallow quantum circuit with a higher number of parameters, meanwhile QAOA offers a deeper ansatz with $2p$ parameters, where $p$ represents the number of layers, so in some cases QAOA could solve a problem using fewer parameters but at the cost of using a deeper circuit. In any case, both have the potential to extract useful results from NISQ devices.

\begin{figure}[!htb]
\begin{center}
\includegraphics[scale=.5]{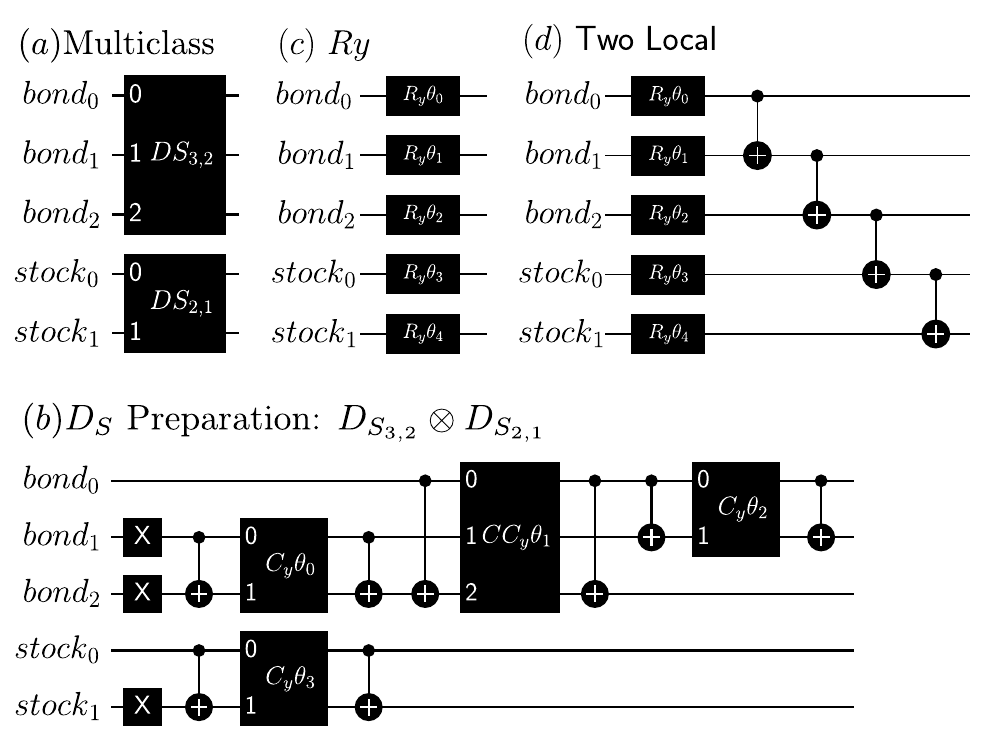}
\end{center}
\caption{Schematic representation of the ansatzes explored in this work: $(a)$ and $(b)$ depict the Dicke State, while $(c)$ and $(d)$ illustrate the $R_y$ and Two Local ansatz, respectively. In this example, the portfolio consists of $5$ assets categorized into $2$ classes—$3$ bonds and $2$ stocks. The optimization goal is to select $2$ bonds and $1$ stock to maximize returns and minimize risks. For a consistent comparison, the different ansatzes are configured to have a comparable number of parameters, with the parameterized Dicke state serving as the reference.}
\label{fig3}
\end{figure}

Some key challenges in applying VQE to practical scenarios are identifying the optimal ansatz, determining efficient parameter initialization methods, and selecting the most suitable classical optimizer. In this study, we systematically explored various ansatzes to determine the most suitable configuration for our specific problem. The investigated ansatzes can be categorized into three distinct types: $(i)$ simple $R_y$ rotation gates, $(ii)$ the extensively studied TwoLocal ansatz along with its variants \cite{buonaiuto2023best}, and $(iii)$ the parameterized Dicke state. A detailed schematic representation of each ansatz is provided in Fig.\ref{fig3}. 

The TwoLocal ansatz is a widely employed variational quantum circuit used in VQE. It consists of two types of parameterized quantum gates arranged in alternating layers: single-qubit rotation gates, such as $R_x$, $R_y$, or $R_z$, and two-qubit entanglement gates, such as $CNOT$. The structure typically begins with rotation gates applied individually to each qubit, followed by layers of entangling gates that couple pairs of qubits, fostering quantum correlations essential for capturing complex solution spaces. This alternating pattern of rotations and entanglement can be repeated multiple times to increase the ansatz's expressivity. Due to its flexibility and relative simplicity, the two-local ansatz has proven effective in approximating solutions for various optimization problems on current NISQ devices.

\section{Portfolio Optimization and Dicke State}
\label{Sec: Dicke}

To address portfolio optimization with quantum computing, first we need to convert Eq.\eqref{mean_variance} to a Quadratic Unconstrained Binary Optimization (QUBO) problem \cite{glover2019quantum}, then the QUBO is converted to an Ising model through the change of variables $x_i = \frac{(1-z_i)}{2}$ \cite{lucas2014ising} 
\begin{eqnarray}
    \min_{x \ \in \ \{0,1\}^n} qx^T\Sigma x - (1-q)x^T\mu + r_f +\lambda\left(Ax-b\right)^2\label{qubo}.
\end{eqnarray}

With that, the equality constraints become a penalty term in the objective function and its intensity is regulated by the Lagrange multiplier $\lambda$. Using information about the structure of the optimization problem, it is possible to encode constraint properties in the ansatz, which produces a quantum state that satisfies the constraints. Therefore, we can remove the penalty term from equation \ref{qubo} setting $\lambda=0$ if the ansatz guarantees a feasible solution.

\begin{table}[h!]
\centering
\begin{tabular}{||c c c c c||} 
 \hline
 Method & Depth CNOTs & $n_{CNOTs}$ & $n_p$ & Topology\\ [0.5ex] 
 \hline\hline
\cite{mukherjee2020preparing} & $O(nk)$ & $5nk-5k^2$ & N/A & all-to-all\\ 
 Ours & $O(nk)$ & $5nk-5k^2$ & $kn - \frac{k(k+1)}{2}$ & all-to-all\\ 
 \cite{bartschi2022short} & $O(k\log\frac{n}{k})$ & $O(nk)$ & N/A & all-to-all\\
\cite{bartschi2022short} & $O(k\sqrt{\frac{n}{k}})$ & $O(nk)$ & N/A & grid\\ 
\cite{wang2024variational} & $2(n-k)$ & $2nk-3k^2$ & $nk-\frac{3k^2}{2}$ & LNN\\
 \cite{wang2024variational} & $2n$ & $nk-\frac{k^2}{2}$ & $\frac{n(k+1)}{2}-\frac{k^2}{4}$ & LNN\\ [1ex]
 \hline
\end{tabular}
\caption{A comparison between different implementations of the Dicke state circuit explored in the literature. The circuits metrics considered are: the complexity of depths of CNOTs, the scaling of number of CNOTs ($n_{CNOTs}$) and number of parameters $n_p$ with $n$ and $k$, and topology. These metrics are relevant for resources optimization for Dicke state preparation on the current noisy quantum devices.}
\label{table:dicke}
\end{table}

To enhance the approach, we can leverage knowledge about the optimization problem structure to design more efficient circuits. For example, let us consider a portfolio optimization problem with $n$ variables and constraints that specify the exact number of $k$ assets that must be selected to minimize risk and maximize return. In realistic scenarios, usually $k \ll n$ in order to obtain simpler and more explainable portfolios with an adequate level of diversification without reducing potential returns. In this situation, feasible candidate solutions must have a Hamming weight equal to $k$. The size of the set of feasible solutions is given by $C_{n,k} = n!/k!(n-k)!$. Consequently, among all $2^n$ quantum states, we eliminate those that do not meet the specified constraints. By initializing the quantum state in a superposition that encompasses only feasible solutions, the search space is effectively reduced from $O(2^n)$ to $O(n^k)$. Hence, the quantum state suitable for handling such constraints is known as the Dicke state \cite{wang2024variational}, which is a quantum state related to a fundamental model of quantum optics that describes the interaction between light and matter \cite{garraway2011dicke}.  The general formula for a uniform distribution of quantum states with $n$ qubits and Hamming weight $k$ is defined by \cite{bartschi2019deterministic}

\begin{eqnarray}
    \vert D_k^n\rangle = C_{n,k}^{-1/2}\sum_{i}\mathcal{P}_i\vert 0 \rangle^{\otimes \ (n-k)}\otimes\vert 1 \rangle^{\otimes \ k}, \label{dicke_state}
\end{eqnarray}
where $\mathcal{P}_i$ represents each possible permutation of a quantum state with $n$ qubits with $k$ qubits equal to $\vert 1 \rangle$. There are several different implementations of the Dicke state (see Table \ref{table:dicke}). The quantum state defined by Eq.\eqref{dicke_state}, was used as the initial state for QAOA in \cite{cook2020quantum, bartschi2020grover, brandhofer2022benchmarking, he2023alignment, niroula2022constrained}, and in these references the authors tested a variety of mixers. Beyond the scope of optimization, Dicke states are relevant to the following fields: quantum game theory \cite{ozdemir2007necessary}, quantum networks \cite{prevedel2009experimental}, quantum metrology \cite{toth2012multipartite}, quantum error correction \cite{ouyang2021permutation} and quantum storage \cite{ouyang2021quantum}.

Beyond their broad applicability, Dicke states also serve as a foundation for our proposed ansatz in variational quantum algorithms. In this study, we assume all-to-all qubit connectivity to emphasize the conceptual contribution of the Dicke-state formulation and its ability to enforce diversification constraints without penalty terms. Nevertheless, the approach is not limited to this topology. As shown in Table~\ref{table:dicke}, existing Dicke-state preparation circuits support all-to-all, grid, and linear-nearest-neighbor (LNN) architectures, with different trade-offs in depth and two-qubit gate count. Hence, the framework is \textit{hardware-agnostic}, as preparing each class subspace only requires initializing a fixed-Hamming-weight superposition. 

This ansatz fits perfectly with the portfolio optimization problem subject to a constraint of a fixed number of products, since it creates a superposition in the space of feasible solutions. To address portfolio optimization through VQE with the Dicke state ansatz, we use its parameterized version (see Fig.\ref{fig3}b): 
\begin{eqnarray}
    \vert D_k^n (\vec{\theta})\rangle = \sum_{i}\mathcal{P}_i a_i(\vec{\theta})\vert 0 \rangle^{\otimes \ (n-k)}\otimes\vert 1 \rangle^{\otimes \ k}, \label{dicke_ansatz}
\end{eqnarray}
where $a_i(\vec{\theta})$ represents amplitude probability as a function of the parameters $\vec{\theta}$. A similar approach was presented by \cite{wang2024variational}, which focused solely on the preparation of a single Dicke state, thus mimicking portfolio optimization without diversification constraints. In this paper, we use the implementation given by \cite{mukherjee2020preparing}, creating a non-uniform Dicke state parameterizing the circuit implementation and incorporating the multiclass optimization by including multiple Dicke states representing the multiple classes. Each Dicke state represents a class with a set of products (number of qubits $n$) and will encode the constraints of the exact number of assets which will be selected by class (parameter $k$) (see Fig.\ref{fig3}a). The equation below dictates the number of parameters $n_p$ of a Dicke state ansatz
\begin{eqnarray}
n_{p} = \sum_{i=1}^{m}k_in_i - \frac{k_i(k_i+1)}{2} \label{reg_params}
\end{eqnarray}
where $k_i$ is the number of states $\vert 1 \rangle$, that corresponds to the Hamming weight associated with the budget constraint. Equation \eqref{reg_params}
was derived empirically, for further details see Suplementary Material.

For $m=1$, we have a unique Dicke state ansatz with $n$ qubits and Hamming weight $k$, but $m>1$ implies a tensor product of different Dicke states. The summation is over the number of classes represented by the number of Dicke states in the tensor product of the initialization. 

\section{Results}
\label{Sec: Results}

We addressed the portfolio optimization problem using the SCIP optimizer, treating its solution as the benchmark reference (see Supplementary Material). The running time was determined by averaging the results from $100$ executions. This average running time served as the basis for comparing the performance of classical methods with hybrid VQE routines. However, it is important to note that we did not expect that the VQE approach would outperform the classical methods in terms of speed.

The experiments were performed in three different scenarios as described in Table \ref{table:scenarios}. For Scenario I, we ran the VQE algorithm for $20$ different ansatzes (Dicke state, $R_y$, and $18$ variations of Two Local). For Scenarios II and III only the Dicke state was used. For all scenarios, we tested $5$ classical optimizers with $1000$ iterations, $100$ randomly sampled initial points (ansatz parameters), $4096$ shots per circuit, totalizing more than $60$ billion executions ($\approx n_{ansatz}\times n_{optimizers} \times n_{executions} \times n_{shots} \times n_{iterations}$) . More details of the variations of the two-local ansatz used here can be found in Supplementary Material. All data used in this work are publicly available and were obtained using the Yahoo Finance API.

\begin{figure}[h!]
\begin{center}
\includegraphics[scale=.55]{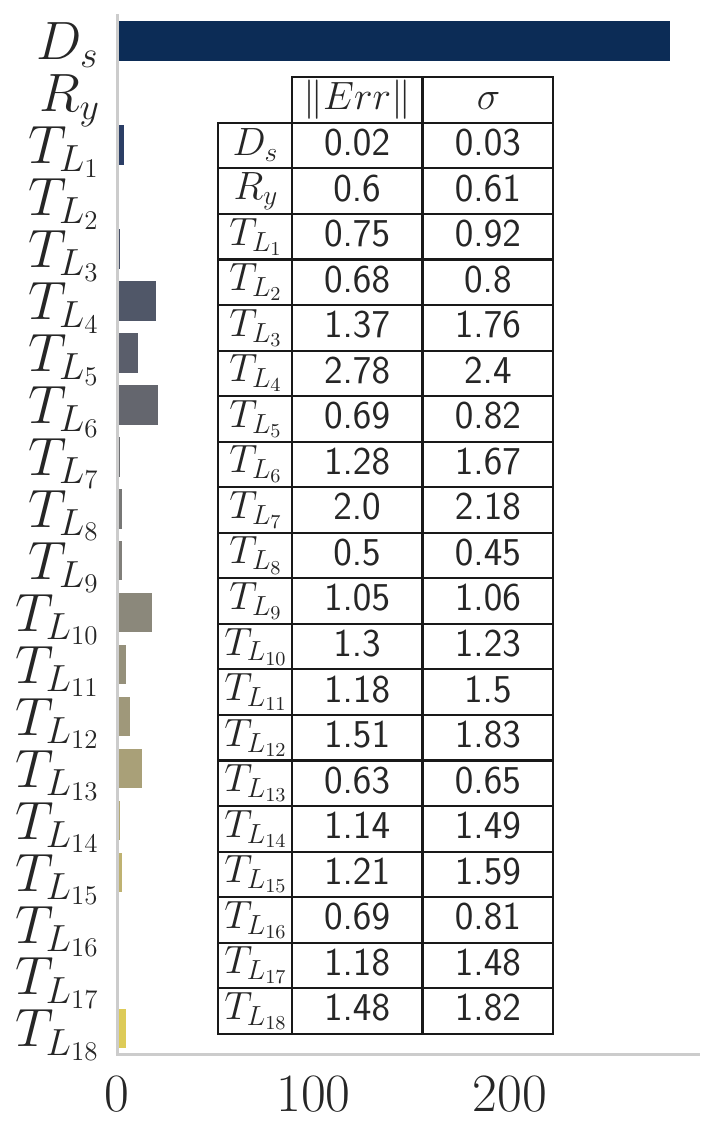}
\end{center}
\caption{Ansatz comparison results for Scenario I taking into account all optimizers. The histogram shows the number of trials where each ansatz found the optimal solution as its most common output. Notably, out of the $20$ ansatzes tested, the Dicke state demonstrated the best performance, independently of the classical optimizer. Out of $500$ runs, in more than $50\%$, the VQE-Dicke state found the optimal result as the one with the highest probability. The inside table shows the absolute error $(\vert\vert Err\vert\vert)$ and the standard deviation $(\sigma)$ between the expected value of the quantum solution and the target. Again, the VQE-Dicke state presented the best metrics.}
\label{fig4_best_ansatz}
\end{figure}

We employ Scenario I to identify the most effective ansatz for multiclass portfolio optimization. Among the $500$ trials conducted per ansatz, the parametrized Dicke state emerged as the best performing approach, regardless of the classical optimizer (see Fig.\ref{fig4_best_ansatz}). For this initial evaluation, we only verified that the optimal solution consistently emerged as the state with the highest probability without considering the magnitude of this probability. It is evident that the Dicke state has been the most effective ansatz so far. Because of that, we only employ it for the other scenarios to assess its performance and evaluate the impact of classical optimizers. 

From a theoretical standpoint, the improved optimization performance of the Dicke-state ansatz arises from its structural design, which confines the variational search to the feasible subspace defined by the diversification constraints. By preparing the quantum state as a tensor product of Dicke states, the search space is reduced from the full $2^n$ Hilbert space to $\prod_i C_{n_i,k_i}$ a substantial reduction in realistic scenarios where $k_i \ll n_i$. This restriction removes infeasible configurations, concentrates the optimization on meaningful portfolio states, and preserves the symmetry of fixed Hamming weight. Consequently, the optimizer operates on a smoother energy landscape, improving convergence stability and accuracy relative to more general ansatzes.
\begin{table}[h!]
\centering
\begin{tabular}{||c c c c c c c||} 
 \hline
 Scenario & $n_{a}$ & $n_{c}$ & $n_{s}$& Ansatz & $n_{p}$ & $n_{\text{search space}}$ \\ [0.5ex] 
 \hline\hline
 I & $10$ & $1$ & $4$ & $\vert D_4^{10}\rangle$ & $30$ &  $210$\\ 
 II & $25$ & $5$ & $5$ & $\vert D_1^{5}\rangle^{\otimes \ 5}$ & $20$ & $3125$ \\ 
 III & $25$ & $5$ & $9$ & $\vert D_2^{5}\rangle^{\otimes \ 2}\vert D_1^{5}\rangle^{\otimes 2}\vert D_3^{5}\rangle$ & $31$ & $25000$\\ 
 \hline
\end{tabular}
\caption{The table shows the different scenarios we use to evaluate the performance of parametrized Dicke state in the context of multiclass portfolio optimization. Where $n_a$ represents the number of assets (that define the number of qubits used in the scenario), $n_c$ number of classes, $n_s$ amount of select assets, $n_p$ number of parameters of Dicke state ansatz. $\vert D_k^n\rangle$ is a superposition of all states with $n$ qubits and $k$ qubits equal $\vert 1\rangle$. Each state represents a class where $n$ is the total number of assets available to choose and $k$ is the number of assets we must select from it. The parameter $\vec{\theta}$ was omitted in the state for better visualization. The last column shows the actual size of search space for each scenario.}
\label{table:scenarios}
\end{table}

\begin{figure}[h!]
\begin{center}
\includegraphics[scale=.65]{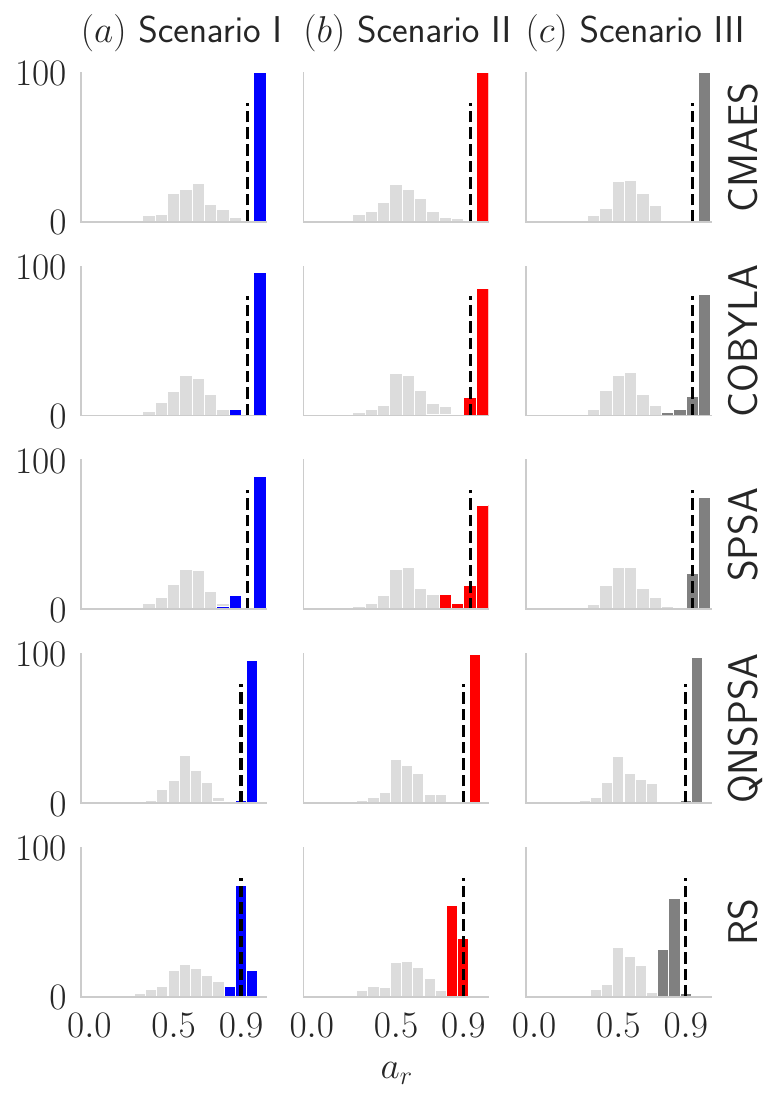}
\end{center}
\caption{Plots $(a)$, $(b)$, and $(c)$ compare the approximation ratio ($a_r$) distributions before (gray) and after (colored) optimization for each scenario (columns) and optimizers (rows). The gray represent $a_r$ distributions with the initial parameters, while the colored ones reflect optimized parameters, based on $100$ experiments per optimizer. A vertical black dashed line marks $a_r = 0.9$. Across all optimizers and scenarios, the initial distributions shift toward $a_r \geq 0.9$, indicating the positive impact of the hybrid approach and optimizers.In Scenario I (blue) all the optimizers guide the solution to the optimal region.}
\label{prob_ar}
\end{figure}

\begin{figure}[!htb]
\begin{center}
\includegraphics[scale=.62]{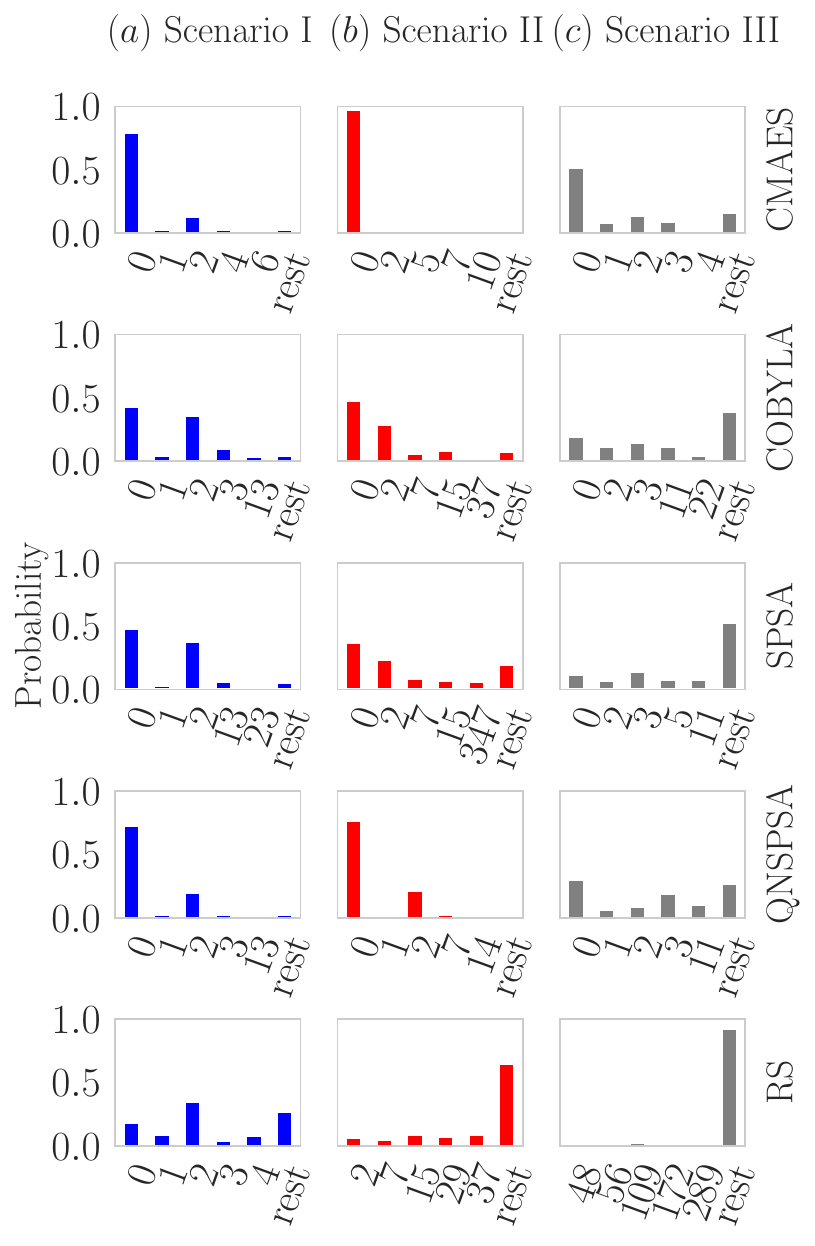}
\end{center}
\caption{Plots $(a)$, $(b)$, and $(c)$ display the sampling probabilities of each bit string, ordered from lower to higher energy, aggregated over 100 runs. The target state is positioned at $0$ based on this energy ordering. We present the five best results from each optimizer across all scenarios, with the tick rest aggregating outcomes outside the top five. Overall, CMA-ES and QNSPSA achieved the best performance, as they concentrate probability distribution in the optimal region and assign the highest probability to the ground state.}
\label{probability}
\end{figure}
 
 Another challenge in the evaluation of hybrid algorithms is to measure the effect of the classical optimizer on the solution. As mentioned before, we tested five different optimizers: CMA-ES \cite{grayver2016exploring}, COBYLA \cite{powell1994direct}, Random Sampler, SPSA \cite{spall1998overview} and QNSPSA \cite{gacon2021simultaneous}. The evaluation takes into account three different metrics: the approximation ratio, the frequency with which the right answer appears in the $100$ trials per optimizer, and the quality of the quantum output for each trial (the probability of measuring the target state). The approximation ratio is defined as \cite{shaydulin2019evaluating, ZhouLeo2020Quantum, herrman2022multi, abbas2024challenges}
\begin{eqnarray}
    a_r = \frac{\left(E_{max}-\langle H \rangle_{\psi(\vec{\theta^*)}}\right)}{(E_{max}-E_0)}, \label{aprox_ratio}
\end{eqnarray}
where $\langle H \rangle_{\psi(\vec{\theta^*)}}$ is the expected value of the problem Hamiltonian computed with the ansatz and the set of parameters obtained at the end of optimization. $E_0$ is the ground-state energy associated with the optimal solution that was calculated by computing the lowest eigenvalue for the problem Hamiltonian. $E_{max}$ represents the highest Hamiltonian eigenvalue. The metric above measures how close or distant the solution is from the optimal. For example, an approximation ratio equal to $1$ means that the solution is equal to the optimal one. 

 When comparing the results of the experiments in different scenarios, it is clear that CMA-ES emerged as the optimizer that exhibited the highest frequency of finding the target state with the highest probability among all the $100$ trials (see Table \ref{table:results}). In terms of time, COBYLA was the fastest optimizer, followed by CMA-ES in second place. QNSPSA, the second solver that demonstrated good performance in finding the global optimal, was surprisingly costly in terms of time, resulting in the worst performance among all optimizers. This result indicates the potential of using CMA-ES as a good choice for hybrid algorithms. More explorations must be done, but, at least for these experiments, CMA-ES surpasses the other optimizers.

\begin{table}[h!]
\centering
\begin{tabular}{|| c c c c c ||}
\hline
 Scenario & Optimizer & All & $p(x^*) \geq  0.95$& Time (s)\\ [0.5ex] 
 \hline\hline
\multirow{5}{*}{I} & CMA-ES  & $\boldsymbol{88}$ & \textcolor{red}{$\bold{25}$} &$82.8 \pm 0.6$ \\
                  & COBYLA & $44$ & $32$ & $29 \pm 4$ \\
                  & QNSPSA & $74$ &  $\boldsymbol{65}$ & $601\pm 3$ \\
                  & RS & $27$ & $0$ &$82.5 \pm 0.7$ \\
                  & SPSA & $49$ & $44$ &$154.5 \pm 0.8$ \\
                  \hline
\multirow{5}{*}{II} & CMA-ES & $\boldsymbol{98}$ & $\boldsymbol{97}$ &$181 \pm 4$ \\
                  & COBYLA & $49$ & $35$ &$43 \pm 6$ \\
                  & QNSPSA & $76$ & $73$ &$981 \pm 9$ \\
                  & RS & $6$ & $0$ &$202 \pm 1$ \\
                  & SPSA & $40$ & $25$ & $302 \pm 6$ \\
                  \hline
\multirow{5}{*}{III} & CMA-ES & $\boldsymbol{70}$ & \textcolor{red}{$\bold{1}$} & $335 \pm 12$ \\
                  & COBYLA & $21$ & $5$ & $119 \pm 21$ \\
                  & QNSPSA & $30$ & $\boldsymbol{27}$ & $1541 \pm 20$\\
                  & RS & $0$ & $0$ & $376 \pm 3$ \\
                  & SPSA & $12$ & $4$ & $467 \pm 17$ \\
                  \hline
\end{tabular}
\caption{A summary of experiments results for each scenario considering only Dicke state ansatz. The column \textit{All} represents the frequency  of the optimal global  solution $(x^*)$, among $100$ experiments, counted if it appears with highest probability. $p(x^*) \geq 0.95$ filter only the quantum outputs which the probability of the optimal state is greater than $0.95$. As can be seen, by applying this filter the frequency in CMAES drastically reduces in some scenarios (numbers indicated in red). The last column represents the average running time and the standard deviation of each optimizer.}
\label{table:results}
\end{table}

 It is clear that the optimizers guide the distribution toward the right direction, this fact can be seen in Fig.\ref{prob_ar}, where we compare the approximation ratio distribution before and after the parameter optimization process. Note that the values of the initial distribution are below $a_r=0.9$, but after optimization we can see a displacement to the right, which means that optimization succeeded in obtaining a set of parameters which generates a quantum state whose state distribution has a high probability of measuring a state with a high approximation ratio.

 Note that even when the probability of measuring the target state is low (see probability of state of measure $0$ in Fig.~\ref{probability}), $a_r \gtrsim 0.9$ for most optimizers, see Fig.~\ref{prob_ar} (colored bars). This phenomenon is linked to Eq.\eqref{aprox_ratio}, which, from a quantum perspective, acts as a weighted average representing the expected value of the Hamiltonian. This implies that the approximation ratio of the quantum output distribution is essentially a weighted sum of the approximation ratios of individual states. As a result, when the system moves toward the optimal region post-optimization, where most states in the distribution have lower energy, the approximation ratio generally becomes higher.
 
The clarity of the results is enhanced upon examining Fig.\ref{probability}. After conducting $100$ experimental runs, we aggregated the data and analyzed the frequency of each bit string, arranging them in ascending order of energy. For improved visualization, we retained only the top five bit strings, consolidating all others into a \textit{rest} category. It is evident that CMA-ES consistently exhibits the highest probability of sampling the optimal solution in all scenarios and dominates the top five regions. QNSPSA, COBYLA, and SPSA show comparable behaviors, ranking second. Although they are capable of achieving the optimal solution, in instances where they do not, they frequently yield solutions that are close to the global optimum.

\begin{figure}[h!]
\begin{center}
\includegraphics[scale=.45]{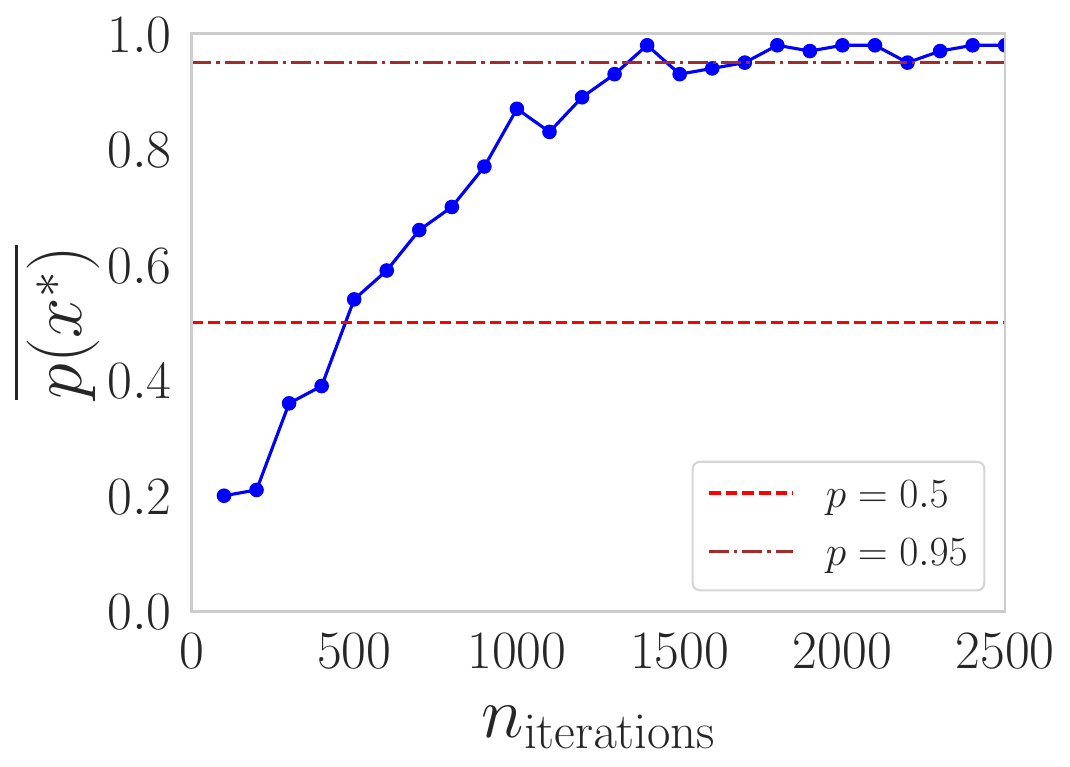}
\end{center}
\caption{The result express how the probability to achieve the target state evolve as we increase the number iterations in CMA-ES optimizer in scenario I. Each point in the curve represents $100$ executions of VQE-Dicke with $4096$ shots for a fixed number of iterations of CMA-ES. Note that, each execution initialize a random set of parameters for VQE. After each execution we extracted the probability of measuring the target state and compute the average among all $100$ runs. As the number of iterations increases, independently of the parameters initialization, $\overline{p(x^*)}$ also increases. After $n_{iterations} \gtrsim 1500$ that the optimal region $p(x^*\geq 0.95)$ is achieved.}
\label{CMA-ES}
\end{figure}

Despite the high performance of CMA-ES in all scenarios, the quality of the quantum output is not consistent. To investigate why $p(x^*)$ is widely distributed, we evaluated the impact of the number of iterations in $p(x^*)$. As can be seen in Fig.\ref{CMA-ES} as the number of iterations increases, the average probability of finding the target state also increases, achieving a more accurate result. The experiment was performed in scenario I and with $n_{interations} \gtrsim 1500$ the solution reaches the optimal region $(p(x^*) \geq 0.95)$. The result strongly suggests that, in addition to the capacity to find the target state, it is possible to improve the quality of the quantum output by increasing the number of iterations.

It is important to mention that recent studies \cite{Scriva2024Challenges, ZhouLeo2020Quantum, maurizio2025genomics} suggest the potential quantum advantage of variational algorithms like VQE and QAOA arises when the number of function evaluations $(n_{calls} \equiv n_{shots} \times n_{iterations})$ remains smaller than the size of the effective search space ($n_{\text{search space}}$). By applying the Dicke state ansatz in our problem, we significantly reduce the search space from $2^n$ by $ \prod_{i}^{n_{c}} C_{n_i, k_i} \sim O(n^k)$. However, for the set of experiments provided here, $n_{calls} \sim 4.096.000  \gg n_{\text{search space}}$ as can be seen in the last column of Table \ref{table:scenarios}.

\section{Conclusion}
\label{Sec: Conclusion}

In this work, we introduce an innovative approach to exploring multiclass portfolio optimization through a parameterized version of multiple Dicke states within a VQE framework. We analyze three distinct scenarios by varying the complexity of the search space and the number of parameters in the parameterized circuit. Our comparative evaluation indicates that Dicke states outperform other ansatzes, making them a suitable choice for this type of problem.

Furthermore, we examine the impact of various optimizers within this hybrid algorithm. Our results indicate that CMA-ES outperforms other optimizers in both execution time and convergence to the optimal solution. However, achieving higher-quality quantum outputs requires a larger number of iterations and tuning of the optimizer parameters. Additionally, all optimizers tested here appear to find parameters that guide the quantum distribution output toward regions close to the ground state. The Random Sampler was shown to be the worst optimizer. 

Another intriguing aspect we began to evaluate is the relationship between state fidelity and optimal parameters, as discussed in the Supplementary Material. In future research, our aim is to address the open questions highlighted here and to further investigate the use of the QAOA with Dicke states for multiclass portfolio optimization. This will include exploring different mixer Hamiltonians.

All our results were obtained through simulations, as our primary objective was to gain a deeper understanding of the algorithm's potential for solving realistic portfolio optimization problems, encoding the diversification constraints in the quantum state preparation. 

Although the proposed Dicke-state framework achieved promising results, further investigation is needed to evaluate its scalability and practical benefits in large-scale, real-world portfolio optimization. The potential advantage of variational algorithms emerges when the number of function evaluations remains smaller than the effective search space, motivating future studies combining the Dicke ansatz with CMA-ES and Conditional Value at Risk (CVaR) to explore possible regimes of quantum advantage. While current simulations assume all-to-all connectivity, the approach is hardware-agnostic and compatible with realistic $2$D or sparse architectures, offering structural efficiency by reducing the feasible subspace and eliminating penalty terms. Preliminary hardware tests on IBM quantum devices can be seen in the Supplementary Material and confirm the expected sensitivity to noise and circuit depth, reinforcing the focus on noiseless simulations to isolate algorithmic behavior. Finally, quantum-optimized portfolios showed consistency with classical allocations in risk–return balance and diversification quality, encouraging future work on explicit financial performance metrics.

Future research is also needed to test these findings on current quantum hardware, comparing the effectiveness of Zero-Noise Extrapolation (ZNE), Probabilistic Error cancellation (PEC) and CVaR in mitigating errors during the calculation of expectation values. Our goal was not to provide definitive proof of this impact, but rather to test the viability of utilizing VQE in a more realistic financial scenario and open a new direction of exploration that is encoding optimization constraints in state preparation. We believe that our findings represent a significant step towards this objective. 

\textbf{Author contribution.} SB and JV conceived the project. SB, JV and GM developed the theoretical framework. JV, GM and VB performed the experiment and collected the data. JV and SB analyzed the experimental data and prepared the figures. All authors contributed to the writing of the manuscript.

\textbf{Disclaimer.} Any opinions, findings, conclusions, or recommendations expressed in this material are those of the authors and do not necessarily reflect the views of Itaú-Unibanco and Institute of Science and Technology of Itaú. This document is not and does not constitute or intend to constitute investment advice or any investment service. It is not and should not be deemed to be an offer to purchase or sell, or a solicitation of an offer to purchase or sell, or a recommendation to purchase or sell any securities or other financial instruments. In addition, all data used in this study comply with the Brazilian General Data Protection Law.

\textbf{Funding declaration.} This research did not receive funding.

\section{Data Availability}

The datasets generated and/or analyzed during the current study are available in the Yahoo Finance and can be downloaded using the following Python library: \href{https://pypi.org/project/yfinance/}{yfinance}. The period of the datasets considered in this work are: 2023-12-18 until 2024-12-16 and 2024-01-02 until 2025-01-02.

\section{Code Availability}

The underlying code for this study is not publicly available but may be made available to qualified researchers on reasonable request from the corresponding author.

\bibliography{ref}

@article{herman2023quantum,
  title={Quantum computing for finance},
  author={Herman, Dylan and Googin, Cody and Liu, Xiaoyuan and Sun, Yue and Galda, Alexey and Safro, Ilya and Pistoia, Marco and Alexeev, Yuri},
  journal={Nature Reviews Physics},
  volume={5},
  number={8},
  pages={450--465},
  year={2023},
  publisher={Nature Publishing Group UK London}
}

@article{glover2019quantum,
  title={Quantum Bridge Analytics I: a tutorial on formulating and using QUBO models},
  author={Glover, Fred and Kochenberger, Gary and Du, Yu},
  journal={4or},
  volume={17},
  number={4},
  pages={335--371},
  year={2019},
  publisher={Springer}
}

@article{lucas2014ising,
  title={Ising formulations of many NP problems},
  author={Lucas, Andrew},
  journal={Frontiers in physics},
  volume={2},
  pages={5},
  year={2014},
  publisher={Frontiers Media SA}
}

@book{griffiths2019introduction,
  title={Introduction to quantum mechanics},
  author={Griffiths, David J and Schroeter, Darrell F},
  year={2019},
  publisher={Cambridge university press}
}

@article{vqe2014,
    author={Peruzzo, Alberto, et al.},
    title={A variational eigenvalue solver on a photonic quantum processor.},
    year={"2014"},
    journal={"Nature communications"},
}

@article{garraway2011dicke,
  title={The Dicke model in quantum optics: Dicke model revisited},
  author={Garraway, Barry M},
  journal={Philosophical Transactions of the Royal Society A: Mathematical, Physical and Engineering Sciences},
  volume={369},
  number={1939},
  pages={1137--1155},
  year={2011},
  publisher={The Royal Society Publishing}
}

@inproceedings{bartschi2019deterministic,
  title={Deterministic preparation of Dicke states},
  author={B{\"a}rtschi, Andreas and Eidenbenz, Stephan},
  booktitle={International Symposium on Fundamentals of Computation Theory},
  pages={126--139},
  year={2019},
  organization={Springer}
}

@article{grayver2016exploring,
  title={Exploring equivalence domain in nonlinear inverse problems using Covariance Matrix Adaption Evolution Strategy (CMAES) and random sampling},
  author={Grayver, Alexander V and Kuvshinov, Alexey V},
  journal={Geophysical Journal International},
  volume={205},
  number={2},
  pages={971--987},
  year={2016},
  publisher={Oxford University Press}
}

@article{spall1998overview,
  title={An overview of the simultaneous perturbation method for efficient optimization},
  author={Spall, James C},
  journal={Johns Hopkins apl technical digest},
  volume={19},
  number={4},
  pages={482--492},
  year={1998}
}

@book{powell1994direct,
  title={A direct search optimization method that models the objective and constraint functions by linear interpolation},
  author={Powell, Michael JD},
  year={1994},
  publisher={Springer}
}

@article{gacon2021simultaneous,
  title={Simultaneous perturbation stochastic approximation of the quantum fisher information},
  author={Gacon, Julien and Zoufal, Christa and Carleo, Giuseppe and Woerner, Stefan},
  journal={Quantum},
  volume={5},
  pages={567},
  year={2021},
  publisher={Verein zur F{\"o}rderung des Open Access Publizierens in den Quantenwissenschaften}
}

@article{javadi2024quantum,
  title={Quantum computing with Qiskit},
  author={Javadi-Abhari, Ali and Treinish, Matthew and Krsulich, Kevin and Wood, Christopher J and Lishman, Jake and Gacon, Julien and Martiel, Simon and Nation, Paul D and Bishop, Lev S and Cross, Andrew W and others},
  journal={arXiv preprint arXiv:2405.08810},
  year={2024}
}

@inproceedings{cook2020quantum,
  title={The quantum alternating operator ansatz on maximum k-vertex cover},
  author={Cook, Jeremy and Eidenbenz, Stephan and B{\"a}rtschi, Andreas},
  booktitle={2020 IEEE International Conference on Quantum Computing and Engineering (QCE)},
  pages={83--92},
  year={2020},
  organization={IEEE}
}

@inproceedings{bartschi2020grover,
  title={Grover mixers for QAOA: Shifting complexity from mixer design to state preparation},
  author={B{\"a}rtschi, Andreas and Eidenbenz, Stephan},
  booktitle={2020 IEEE International Conference on Quantum Computing and Engineering (QCE)},
  pages={72--82},
  year={2020},
  organization={IEEE}
}

@article{brandhofer2022benchmarking,
  title={Benchmarking the performance of portfolio optimization with QAOA},
  author={Brandhofer, Sebastian and Braun, Daniel and Dehn, Vanessa and Hellstern, Gerhard and H{\"u}ls, Matthias and Ji, Yanjun and Polian, Ilia and Bhatia, Amandeep Singh and Wellens, Thomas},
  journal={Quantum Information Processing},
  volume={22},
  number={1},
  pages={25},
  year={2022},
  publisher={Springer}
}

@article{he2023alignment,
  title={Alignment between initial state and mixer improves QAOA performance for constrained optimization},
  author={He, Zichang and Shaydulin, Ruslan and Chakrabarti, Shouvanik and Herman, Dylan and Li, Changhao and Sun, Yue and Pistoia, Marco},
  journal={npj Quantum Information},
  volume={9},
  number={1},
  pages={121},
  year={2023},
  publisher={Nature Publishing Group UK London}
}

@article{niroula2022constrained,
  title={Constrained quantum optimization for extractive summarization on a trapped-ion quantum computer},
  author={Niroula, Pradeep and Shaydulin, Ruslan and Yalovetzky, Romina and Minssen, Pierre and Herman, Dylan and Hu, Shaohan and Pistoia, Marco},
  journal={Scientific Reports},
  volume={12},
  number={1},
  pages={17171},
  year={2022},
  publisher={Nature Publishing Group UK London}
}

@article{barkoutsos2020improving,
  title={Improving variational quantum optimization using CVaR},
  author={Barkoutsos, Panagiotis Kl and Nannicini, Giacomo and Robert, Anton and Tavernelli, Ivano and Woerner, Stefan},
  journal={Quantum},
  volume={4},
  pages={256},
  year={2020},
  publisher={Verein zur F{\"o}rderung des Open Access Publizierens in den Quantenwissenschaften}
}

@article{barron2024provable,
  title={Provable bounds for noise-free expectation values computed from noisy samples},
  author={Barron, Samantha V and Egger, Daniel J and Pelofske, Elijah and B{\"a}rtschi, Andreas and Eidenbenz, Stephan and Lehmkuehler, Matthis and Woerner, Stefan},
  journal={Nature Computational Science},
  pages={1--11},
  year={2024},
  publisher={Nature Publishing Group US New York}
}

@article{mukherjee2020preparing,
  title={Preparing Dicke states on a quantum computer},
  author={Mukherjee, Chandra Sekhar and Maitra, Subhamoy and Gaurav, Vineet and Roy, Dibyendu},
  journal={IEEE Transactions on Quantum Engineering},
  volume={1},
  pages={1--17},
  year={2020},
  publisher={IEEE}
}

@article{wang2024variational,
  title={Variational quantum eigensolver with linear depth problem-inspired ansatz for solving portfolio optimization in finance},
  author={Wang, Shengbin and Wang, Peng and Li, Guihui and Zhao, Shubin and Zhao, Dongyi and Wang, Jing and Fang, Yuan and Dou, Menghan and Gu, Yongjian and Wu, Yu-Chun and others},
  journal={arXiv preprint arXiv:2403.04296},
  year={2024}
}

@article{kandala2018extending,
  title={Extending the computational reach of a noisy superconducting quantum processor},
  author={Kandala, Abhinav and Temme, Kristan and Corcoles, Antonio D and Mezzacapo, Antonio and Chow, Jerry M and Gambetta, Jay M},
  journal={arXiv preprint arXiv:1805.04492},
  year={2018}
}

@inproceedings{giurgica2020digital,
  title={Digital zero noise extrapolation for quantum error mitigation},
  author={Giurgica-Tiron, Tudor and Hindy, Yousef and LaRose, Ryan and Mari, Andrea and Zeng, William J},
  booktitle={2020 IEEE International Conference on Quantum Computing and Engineering (QCE)},
  pages={306--316},
  year={2020},
  organization={IEEE}
}

@article{kim2023scalable,
  title={Scalable error mitigation for noisy quantum circuits produces competitive expectation values},
  author={Kim, Youngseok and Wood, Christopher J and Yoder, Theodore J and Merkel, Seth T and Gambetta, Jay M and Temme, Kristan and Kandala, Abhinav},
  journal={Nature Physics},
  volume={19},
  number={5},
  pages={752--759},
  year={2023},
  publisher={Nature Publishing Group UK London}
}

@article{cerezo2021variational,
  title={Variational quantum algorithms},
  author={Cerezo, Marco and Arrasmith, Andrew and Babbush, Ryan and Benjamin, Simon C and Endo, Suguru and Fujii, Keisuke and McClean, Jarrod R and Mitarai, Kosuke and Yuan, Xiao and Cincio, Lukasz and others},
  journal={Nature Reviews Physics},
  volume={3},
  number={9},
  pages={625--644},
  year={2021},
  publisher={Nature Publishing Group UK London}
}

@article{abbas2024challenges,
  title={Challenges and opportunities in quantum optimization},
  author={Abbas, Amira and Ambainis, Andris and Augustino, Brandon and B{\"a}rtschi, Andreas and Buhrman, Harry and Coffrin, Carleton and Cortiana, Giorgio and Dunjko, Vedran and Egger, Daniel J and Elmegreen, Bruce G and others},
  journal={Nature Reviews Physics},
  pages={1--18},
  year={2024},
  publisher={Nature Publishing Group UK London}
}

@article{farhi2014quantum,
  title={A quantum approximate optimization algorithm},
  author={Farhi, Edward and Goldstone, Jeffrey and Gutmann, Sam},
  journal={arXiv preprint arXiv:1411.4028},
  year={2014}
}

@article{biamonte2017quantum,
  title={Quantum machine learning},
  author={Biamonte, Jacob and Wittek, Peter and Pancotti, Nicola and Rebentrost, Patrick and Wiebe, Nathan and Lloyd, Seth},
  journal={Nature},
  volume={549},
  number={7671},
  pages={195--202},
  year={2017},
  publisher={Nature Publishing Group UK London}
}

@article{mcardle2020quantum,
  title={Quantum computational chemistry},
  author={McArdle, Sam and Endo, Suguru and Aspuru-Guzik, Al{\'a}n and Benjamin, Simon C and Yuan, Xiao},
  journal={Reviews of Modern Physics},
  volume={92},
  number={1},
  pages={015003},
  year={2020},
  publisher={APS}
}

@article{bauer2020quantum,
  title={Quantum algorithms for quantum chemistry and quantum materials science},
  author={Bauer, Bela and Bravyi, Sergey and Motta, Mario and Chan, Garnet Kin-Lic},
  journal={Chemical reviews},
  volume={120},
  number={22},
  pages={12685--12717},
  year={2020},
  publisher={ACS Publications}
}

@article{preskill2018quantum,
  title={Quantum computing in the NISQ era and beyond},
  author={Preskill, John},
  journal={Quantum},
  volume={2},
  pages={79},
  year={2018},
  publisher={Verein zur F{\"o}rderung des Open Access Publizierens in den Quantenwissenschaften}
}

@article{wang2021noise,
  title={Noise-induced barren plateaus in variational quantum algorithms},
  author={Wang, Samson and Fontana, Enrico and Cerezo, Marco and Sharma, Kunal and Sone, Akira and Cincio, Lukasz and Coles, Patrick J},
  journal={Nature communications},
  volume={12},
  number={1},
  pages={6961},
  year={2021},
  publisher={Nature Publishing Group UK London}
}

@book{nielsen2010quantum,
  title={Quantum computation and quantum information},
  author={Nielsen, Michael A and Chuang, Isaac L},
  year={2010},
  publisher={Cambridge university press}
}

@article{van2023probabilistic,
  title={Probabilistic error cancellation with sparse Pauli--Lindblad models on noisy quantum processors},
  author={Van Den Berg, Ewout and Minev, Zlatko K and Kandala, Abhinav and Temme, Kristan},
  journal={Nature physics},
  volume={19},
  number={8},
  pages={1116--1121},
  year={2023},
  publisher={Nature Publishing Group UK London}
}

@article{gupta2024probabilistic,
  title={Probabilistic error cancellation for dynamic quantum circuits},
  author={Gupta, Riddhi S and Van Den Berg, Ewout and Takita, Maika and Riste, Diego and Temme, Kristan and Kandala, Abhinav},
  journal={Physical Review A},
  volume={109},
  number={6},
  pages={062617},
  year={2024},
  publisher={APS}
}

@article{filippov2023scalable,
  title={Scalable tensor-network error mitigation for near-term quantum computing},
  author={Filippov, Sergei and Leahy, Matea and Rossi, Matteo AC and Garc{\'\i}a-P{\'e}rez, Guillermo},
  journal={arXiv preprint arXiv:2307.11740},
  year={2023}
}

@article{kim2023evidence,
  title={Evidence for the utility of quantum computing before fault tolerance},
  author={Kim, Youngseok and Eddins, Andrew and Anand, Sajant and Wei, Ken Xuan and Van Den Berg, Ewout and Rosenblatt, Sami and Nayfeh, Hasan and Wu, Yantao and Zaletel, Michael and Temme, Kristan and others},
  journal={Nature},
  volume={618},
  number={7965},
  pages={500--505},
  year={2023},
  publisher={Nature Publishing Group UK London}
}

@article{van2022model,
  title={Model-free readout-error mitigation for quantum expectation values},
  author={Van Den Berg, Ewout and Minev, Zlatko K and Temme, Kristan},
  journal={Physical Review A},
  volume={105},
  number={3},
  pages={032620},
  year={2022},
  publisher={APS}
}

@article{nation2021scalable,
  title={Scalable mitigation of measurement errors on quantum computers},
  author={Nation, Paul D and Kang, Hwajung and Sundaresan, Neereja and Gambetta, Jay M},
  journal={PRX Quantum},
  volume={2},
  number={4},
  pages={040326},
  year={2021},
  publisher={APS}
}

@book{weinberg2015lectures,
  title={Lectures on quantum mechanics},
  author={Weinberg, Steven},
  year={2015},
  publisher={Cambridge University Press}
}

@article{chen2024validating,
  title={Validating Large-Scale Quantum Machine Learning: Efficient Simulation of Quantum Support Vector Machines Using Tensor Networks},
  author={Chen, Kuan-Cheng and Li, Tai-Yue and Wang, Yun-Yuan and See, Simon and Wang, Chun-Chieh and Wille, Robert and Chen, Nan-Yow and Yang, An-Cheng and Lin, Chun-Yu},
  journal={Machine Learning: Science and Technology},
  year={2024}
}

@article{smith1999neural,
  title={Neural networks for combinatorial optimization: a review of more than a decade of research},
  author={Smith, Kate A},
  journal={Informs journal on Computing},
  volume={11},
  number={1},
  pages={15--34},
  year={1999},
  publisher={INFORMS}
}

@inproceedings{prates2019learning,
  title={Learning to solve np-complete problems: A graph neural network for decision tsp},
  author={Prates, Marcelo and Avelar, Pedro HC and Lemos, Henrique and Lamb, Luis C and Vardi, Moshe Y},
  booktitle={Proceedings of the AAAI conference on artificial intelligence},
  volume={33},
  number={01},
  pages={4731--4738},
  year={2019}
}

@inproceedings{bartschi2022short,
  title={Short-depth circuits for Dicke state preparation},
  author={B{\"a}rtschi, Andreas and Eidenbenz, Stephan},
  booktitle={2022 IEEE International Conference on Quantum Computing and Engineering (QCE)},
  pages={87--96},
  year={2022},
  organization={IEEE}
}

@book{junger200950,
  title={50 Years of integer programming 1958-2008: From the early years to the state-of-the-art},
  author={J{\"u}nger, Michael and Liebling, Thomas M and Naddef, Denis and Nemhauser, George L and Pulleyblank, William R and Reinelt, Gerhard and Rinaldi, Giovanni and Wolsey, Laurence A},
  year={2009},
  publisher={Springer Science \& Business Media}
}

@book{gonzalez2007handbook,
  title={Handbook of approximation algorithms and metaheuristics},
  author={Gonzalez, Teofilo F},
  year={2007},
  publisher={Chapman and Hall/CRC}
}

@article{markowitz1952portfolio,
  title={Portfolio selection},
  author={Markowitz, Harry M},
  journal={Journal of finance},
  volume={7},
  number={1},
  pages={71--91},
  year={1952}
}

@article{ozdemir2007necessary,
  title={A necessary and sufficient condition to play games in quantum mechanical settings},
  author={{\"O}zdemir, Sahin K and Shimamura, Junichi and Imoto, Nobuyuki},
  journal={New Journal of Physics},
  volume={9},
  number={2},
  pages={43},
  year={2007},
  publisher={IOP Publishing}
}

@article{prevedel2009experimental,
  title={Experimental realization of Dicke states of up to six qubits for multiparty quantum networking},
  author={Prevedel, Robert and Cronenberg, Gunther and Tame, Mark S and Paternostro, Mauro and Walther, Philip and Kim, Mu-Seong and Zeilinger, Anton},
  journal={Physical review letters},
  volume={103},
  number={2},
  pages={020503},
  year={2009},
  publisher={APS}
}

@article{toth2012multipartite,
  title={Multipartite entanglement and high-precision metrology},
  author={T{\'o}th, G{\'e}za},
  journal={Physical Review A—Atomic, Molecular, and Optical Physics},
  volume={85},
  number={2},
  pages={022322},
  year={2012},
  publisher={APS}
}

@inproceedings{ouyang2021permutation,
  title={Permutation-invariant quantum coding for quantum deletion channels},
  author={Ouyang, Yingkai},
  booktitle={2021 IEEE International Symposium on Information Theory (ISIT)},
  pages={1499--1503},
  year={2021},
  organization={IEEE}
}

@article{ouyang2021quantum,
  title={Quantum storage in quantum ferromagnets},
  author={Ouyang, Yingkai},
  journal={Physical Review B},
  volume={103},
  number={14},
  pages={144417},
  year={2021},
  publisher={APS}
}

@article{cerezo2022challenges,
  title={Challenges and opportunities in quantum machine learning},
  author={Cerezo, Marco and Verdon, Guillaume and Huang, Hsin-Yuan and Cincio, Lukasz and Coles, Patrick J},
  journal={Nature computational science},
  volume={2},
  number={9},
  pages={567--576},
  year={2022},
  publisher={Nature Publishing Group US New York}
}

@article{liu2018quantum,
  title={Quantum machine learning for quantum anomaly detection},
  author={Liu, Nana and Rebentrost, Patrick},
  journal={Physical Review A},
  volume={97},
  number={4},
  pages={042315},
  year={2018},
  publisher={APS}
}

@article{jerbi2023quantum,
  title={Quantum machine learning beyond kernel methods},
  author={Jerbi, Sofiene and Fiderer, Lukas J and Poulsen Nautrup, Hendrik and K{\"u}bler, Jonas M and Briegel, Hans J and Dunjko, Vedran},
  journal={Nature Communications},
  volume={14},
  number={1},
  pages={517},
  year={2023},
  publisher={Nature Publishing Group UK London}
}

@article{schuld2019quantum,
  title={Quantum machine learning in feature Hilbert spaces},
  author={Schuld, Maria and Killoran, Nathan},
  journal={Physical review letters},
  volume={122},
  number={4},
  pages={040504},
  year={2019},
  publisher={APS}
}

@article{von2021quantum,
  title={Quantum computing enhanced computational catalysis},
  author={von Burg, Vera and Low, Guang Hao and H{\"a}ner, Thomas and Steiger, Damian S and Reiher, Markus and Roetteler, Martin and Troyer, Matthias},
  journal={Physical Review Research},
  volume={3},
  number={3},
  pages={033055},
  year={2021},
  publisher={APS}
}

@article{egger2020quantum,
  title={Quantum computing for finance: State-of-the-art and future prospects},
  author={Egger, Daniel J and Gambella, Claudio and Marecek, Jakub and McFaddin, Scott and Mevissen, Martin and Raymond, Rudy and Simonetto, Andrea and Woerner, Stefan and Yndurain, Elena},
  journal={IEEE Transactions on Quantum Engineering},
  volume={1},
  pages={1--24},
  year={2020},
  publisher={IEEE}
}

@article{ramos2021quantum,
  title={Quantum unary approach to option pricing},
  author={Ramos-Calderer, Sergi and P{\'e}rez-Salinas, Adri{\'a}n and Garc{\'\i}a-Mart{\'\i}n, Diego and Bravo-Prieto, Carlos and Cortada, Jorge and Planaguma, Jordi and Latorre, Jos{\'e} I},
  journal={Physical Review A},
  volume={103},
  number={3},
  pages={032414},
  year={2021},
  publisher={APS}
}

@article{wilkens2023quantum,
  title={Quantum computing for financial risk measurement},
  author={Wilkens, Sascha and Moorhouse, Joe},
  journal={Quantum Information Processing},
  volume={22},
  number={1},
  pages={51},
  year={2023},
  publisher={Springer}
}

@article{naik2025portfolio,
  title={From portfolio optimization to quantum blockchain and security: A systematic review of quantum computing in finance},
  author={Naik, Abha Satyavan and Yeniaras, Esra and Hellstern, Gerhard and Prasad, Grishma and Vishwakarma, Sanjay Kumar Lalta Prasad},
  journal={Financial Innovation},
  volume={11},
  number={1},
  pages={1--67},
  year={2025},
  publisher={Springer}
}

@article{buonaiuto2023best,
  title={Best practices for portfolio optimization by quantum computing, experimented on real quantum devices},
  author={Buonaiuto, Giuseppe and Gargiulo, Francesco and De Pietro, Giuseppe and Esposito, Massimo and Pota, Marco},
  journal={Scientific Reports},
  volume={13},
  number={1},
  pages={19434},
  year={2023},
  publisher={Nature Publishing Group UK London}
}

@article{mugel2022dynamic,
  title={Dynamic portfolio optimization with real datasets using quantum processors and quantum-inspired tensor networks},
  author={Mugel, Samuel and Kuchkovsky, Carlos and S{\'a}nchez, Escol{\'a}stico and Fern{\'a}ndez-Lorenzo, Samuel and Luis-Hita, Jorge and Lizaso, Enrique and Or{\'u}s, Rom{\'a}n},
  journal={Physical Review Research},
  volume={4},
  number={1},
  pages={013006},
  year={2022},
  publisher={APS}
}

@article{cong2019quantum,
  title={Quantum convolutional neural networks},
  author={Cong, Iris and Choi, Soonwon and Lukin, Mikhail D},
  journal={Nature Physics},
  volume={15},
  number={12},
  pages={1273--1278},
  year={2019},
  publisher={Nature Publishing Group UK London}
}

@article{jarret2018improved,
  title={Improved quantum backtracking algorithms using effective resistance estimates},
  author={Jarret, Michael and Wan, Kianna},
  journal={Physical Review A},
  volume={97},
  number={2},
  pages={022337},
  year={2018},
  publisher={APS}
}

@article{campbell2019applying,
  title={Applying quantum algorithms to constraint satisfaction problems},
  author={Campbell, Earl and Khurana, Ankur and Montanaro, Ashley},
  journal={Quantum},
  volume={3},
  pages={167},
  year={2019},
  publisher={Verein zur F{\"o}rderung des Open Access Publizierens in den Quantenwissenschaften}
}

@article{montanaro2020quantum,
  title={Quantum speedup of branch-and-bound algorithms},
  author={Montanaro, Ashley},
  journal={Physical Review Research},
  volume={2},
  number={1},
  pages={013056},
  year={2020},
  publisher={APS}
}

@article{finvzgar2024quantum,
  title={Quantum-informed recursive optimization algorithms},
  author={Fin{\v{z}}gar, Jernej Rudi and Kerschbaumer, Aron and Schuetz, Martin JA and Mendl, Christian B and Katzgraber, Helmut G},
  journal={PRX Quantum},
  volume={5},
  number={2},
  pages={020327},
  year={2024},
  publisher={APS}
}

@article{egger2021warm,
  title={Warm-starting quantum optimization},
  author={Egger, Daniel J and Mare{\v{c}}ek, Jakub and Woerner, Stefan},
  journal={Quantum},
  volume={5},
  pages={479},
  year={2021},
  publisher={Verein zur F{\"o}rderung des Open Access Publizierens in den Quantenwissenschaften}
}

@article{magann2022feedback,
  title={Feedback-based quantum optimization},
  author={Magann, Alicia B and Rudinger, Kenneth M and Grace, Matthew D and Sarovar, Mohan},
  journal={Physical Review Letters},
  volume={129},
  number={25},
  pages={250502},
  year={2022},
  publisher={APS}
}

@article{junger1995traveling,
  title={The traveling salesman problem},
  author={J{\"u}nger, Michael and Reinelt, Gerhard and Rinaldi, Giovanni},
  journal={Handbooks in operations research and management science},
  volume={7},
  pages={225--330},
  year={1995},
  publisher={Elsevier}
}

@book{toth2002vehicle,
  title={The vehicle routing problem},
  author={Toth, Paolo and Vigo, Daniele},
  year={2002},
  publisher={SIAM}
}

@article{martello2000three,
  title={The three-dimensional bin packing problem},
  author={Martello, Silvano and Pisinger, David and Vigo, Daniele},
  journal={Operations research},
  volume={48},
  number={2},
  pages={256--267},
  year={2000},
  publisher={INFORMS}
}

@book{mansini2015linear,
  title={Linear and mixed integer programming for portfolio optimization},
  author={Mansini, Renata and W{\'L} ‚odzimierz Ogryczak and Speranza, M Grazia and EURO: The Association of European Operational Research Societies},
  volume={21},
  year={2015},
  publisher={Springer}
}

@article{fanizza2020beyond,
  title={Beyond the swap test: optimal estimation of quantum state overlap},
  author={Fanizza, Marco and Rosati, Matteo and Skotiniotis, Michalis and Calsamiglia, John and Giovannetti, Vittorio},
  journal={Physical review letters},
  volume={124},
  number={6},
  pages={060503},
  year={2020},
  publisher={APS}
}

@article{herrman2022multi,
  title={Multi-angle quantum approximate optimization algorithm},
  author={Herrman, Rebekah and Lotshaw, Phillip C and Ostrowski, James and Humble, Travis S and Siopsis, George},
  journal={Scientific Reports},
  volume={12},
  number={1},
  pages={6781},
  year={2022},
  publisher={Nature Publishing Group UK London}
}

@inproceedings{shaydulin2019evaluating,
  title={Evaluating quantum approximate optimization algorithm: A case study},
  author={Shaydulin, Ruslan and Alexeev, Yuri},
  booktitle={2019 tenth international green and sustainable computing conference (IGSC)},
  pages={1--6},
  year={2019},
  organization={IEEE}
}

@article{ZhouLeo2020Quantum,
  title = {Quantum Approximate Optimization Algorithm: Performance, Mechanism, and Implementation on Near-Term Devices},
  author = {Zhou, Leo and Wang, Sheng-Tao and Choi, Soonwon and Pichler, Hannes and Lukin, Mikhail D.},
  journal = {Phys. Rev. X},
  volume = {10},
  issue = {2},
  pages = {021067},
  numpages = {23},
  year = {2020},
  month = {Jun},
  publisher = {American Physical Society},
  doi = {},
  url = {}}

@article{Scriva2024Challenges,
  title = {Challenges of variational quantum optimization with measurement shot noise},
  author = {Scriva, Giuseppe and Astrakhantsev, Nikita and Pilati, Sebastiano and Mazzola, Guglielmo},
  journal = {Phys. Rev. A},
  volume = {109},
  issue = {3},
  pages = {032408},
  numpages = {14},
  year = {2024},
  month = {Mar},
  publisher = {American Physical Society},
  doi = {},
  url ={}}

@misc{maurizio2025genomics,
      title={Quantum computing for genomics: conceptual challenges and practical perspectives}, 
      author={Aurora Maurizio and Guglielmo Mazzola},
      year={2025},
      eprint={},
      archivePrefix={arXiv},
      primaryClass={quant-ph},
      url= {}}

@article{Thakkar2024,
	author = {Thakkar, Sohum and Kazdaghli, Skander and Mathur, Natansh and Kerenidis, Iordanis and Ferreira--Martins, Andr{\'e}J. and Brito, Samurai},
	date = {2024/05/07},
	doi = {},
	id = {Thakkar2024},
	isbn = {2524-4914},
	journal = {Quantum Machine Intelligence},
	number = {1},
	pages = {27},
	title = {Improved financial forecasting via quantum machine learning},
	url = {},
	volume = {6},
	year = {2024}}

\newpage
\maketitle
\onecolumngrid

\section*{Supplementary Methods}

\section{Experiment settings}
\label{appendixA}

The VQE algorithm simulations were performed using the instance ml.g5.xlarge on Amazon SageMaker. This machine has four virtual CPUs, 16 GB RAM and NVIDIA GPU A10G with 24 GB GDDR6 VRAM. This system runs on Linux, with Python version 3.9.5, CUDA version 12.4, NVIDIA Graphics driver version 550.144.03. The Python libraries used to perform the simulations are: qiskit version 1.3.2 \cite{javadi2024quantum}, qiskit-aer-gpu version 0.15.1, qiskit-algorithms version 0.3.1, qiskit-optimization version 0.6.1, numpy version 2.0.2, optuna version 4.1.0 and CMA-ES version 0.11.1.
All initial points were generated with NumPy random methods using the seed $42$, with the exception for CMA-ES and Random Sampler due to optuna's parameter sampling method. However, we guarantee that all parameters are in the interval $[0, 2\pi]$, independently of the method used to sample the parameters values.
CMA-ES and Random Sampler were used through the Optuna interface, whereas COBYLA, SPSA and QNSPSA were used through the Qiskit algorithms package.

As mentioned in the main paper, we solved each scenario proposed with the classical optimizer SCIP and considered its results as a reference. These results are summarized in Table \ref{table:scip}.
\begin{table}[h!]
\centering
\begin{tabular}{||c c c||} 
 \hline
 Scenario & Objective function value & Running time (s) \\ [0.5ex] 
 \hline\hline
 I & -0.818106 & 0.01 $\pm$ 0.001\\ 
 II & -1.474237 & 0.03 $\pm$ 0.001\\
 III & -2.00332 & 0.03 $\pm$ 0.001\\[1ex] 
 \hline
\end{tabular}
\caption{A summary of the reference values computed using SCIP optimizer for each scenario. The column Objective function value holds the optimal values for each case. The column Running time represents the mean time spent and the standard deviation of 100 runs of the optimization problem.}
\label{table:scip}
\end{table}

\section{TwoLocal ansatz}
\label{twolocal}

The TwoLocal circuit is an ansatz that alternates rotation and entanglement layers. The rotation gates are applied in each qubit individually and the gates $RX$, $RY$, $RZ$, $U_1$, $U_2$ and $U_3$ are some examples of rotation gates that we can use in this ansatz.  Meanwhile, the entanglement gates are used to create entanglement between pairs of qubits considered in each entanglement structure design, for instance: we can create entanglement between all qubits or just in some qubits in a way that connects the qubits through a line. In order to create the desired entanglement structure, we can use gates such as: $CX$, $CY$, $CZ$, $CRX$, $CRY$ and $CRZ$. For instance, in Table \ref{table:twolocal} we show some TwoLocal settings used in Scenario I experiments.

\begin{table*}[!htb]
\centering
\begin{tabular}{||c c c c c c c c ||} 
 \hline
 Name & Qubits & Parameters & Rotation Blocks & Entanglement Blocks & Entanglement & Repetitions & Skip Final Rotation Layer \\ [0.5ex] 
 \hline\hline
 TwoLocal1 & 10 & 30 & 3 x RY & CX & Full & 1 & True\\
 TwoLocal2 & 10 & 30 & 1 x RY & CX & Full & 2 & True\\
 TwoLocal3 & 10 & 30 & 1 x RY & CX & Full & 3 & False\\
 TwoLocal4 & 10 & 30 & 3 x RY & CX & Linear & 1 & True\\
 TwoLocal5 & 10 & 30 & 1 x RY & CX & Linear & 2 & True\\
 TwoLocal6 & 10 & 30 & 1 x RY & CX & Linear & 3 & False\\
 TwoLocal7 & 10 & 30 & 3 x RY & CX & Reverse Linear & 1 & True\\
 TwoLocal8 & 10 & 30 & 1 x RY & CX & Reverse Linear & 2 & True\\
 TwoLocal9 & 10 & 30 & 1 x RY & CX & Reverse Linear & 3 & False\\
 TwoLocal10 & 10 & 30 & 3 x RY & CX & Circular & 1 & True\\
 TwoLocal11 & 10 & 30 & 1 x RY & CX & Circular & 2 & True\\
 TwoLocal12 & 10 & 30 & 1 x RY & CX & Circular & 3 & False\\
 TwoLocal13 & 10 & 30 & 3 x RY & CX & SCA & 1 & True\\
 TwoLocal14 & 10 & 30 & 1 x RY & CX & SCA & 2 & True\\
 TwoLocal15 & 10 & 30 & 1 x RY & CX & SCA & 3 & False\\
 TwoLocal16 & 10 & 30 & 3 x RY & CX & Pairwise & 1 & True\\
 TwoLocal17 & 10 & 30 & 1 x RY & CX & Pairwise & 2 & True\\
 TwoLocal18 & 10 & 30 & 1 x RY & CX & Pairwise & 3 & False\\ [1ex] 
 \hline
\end{tabular}
\caption{Settings used to compose each TwoLocal variant consired in the experiments in Scenario I. All circuits for Scenario I were created to have 30 parameters and 10 qubits, then we just varied the amount of rotation blocks, entanglement structure, the number of layers repetitions and whether we skip the final rotation layer. In this case we only used $RY$ and $CX$ gates because relative phases are not relevant.}
\label{table:twolocal}
\end{table*}

\section{Derivation of the number of parameters in the parametrized Dicke State}
\label{num_parameters}

\begin{figure}[!htb]
\begin{center}
\includegraphics[scale=.6]{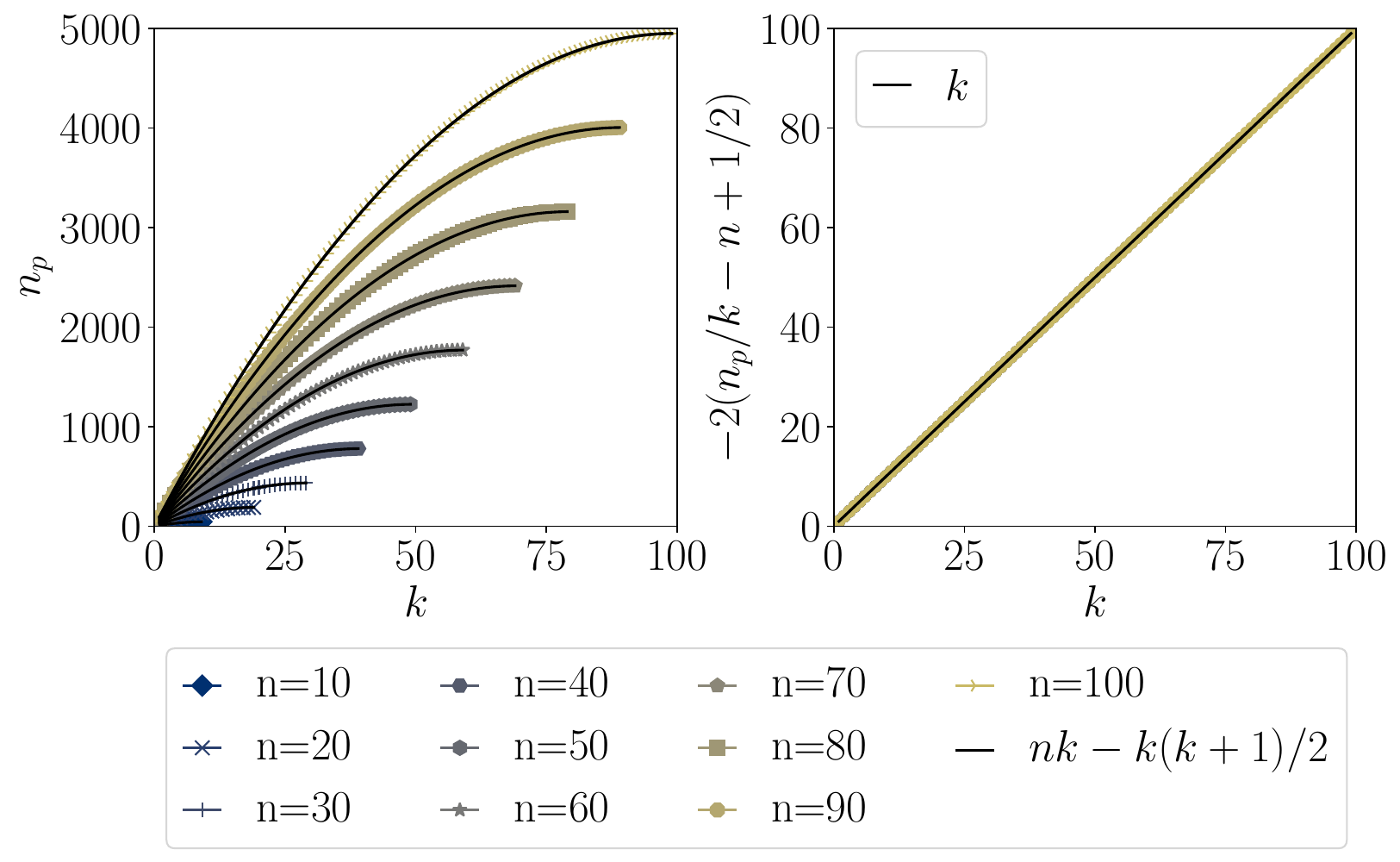}
\end{center}
\caption{Scaling of the number of parameters in Dicke state ansatz for different numbers of qubits and varying the Hamming weight $k$. Note that for $k=0$ and $k=n$ we don't need a parameterized Dicke state, because $k=0$ is equivalent to $\vert0\rangle^{\otimes n}$ and $k=n$ is the state $\vert1\rangle^{\otimes n}$, then the range applied to $k$ is $[1, n-1]$. The curves defined by different values of $n$ and $k$ can be unified by the expression $nk-k(k+1)/2$, as depicted in the plot in the right. }
\label{fig_parameters}
\end{figure}

The total number of parameters is directly related to the number of gates $CRY$ and $CCRY$, and each of these gates has one free parameter. We defined a range of values for $n$ and considered $[1, n-1]$ as the range for $k$. Then, we constructed the Dicke state ansatz for all values of $n$ and $k$ defined in the ranges mentioned above and computed the total number of parameterized gates. With these data in our hands, we were able to plot the curves (see Figure \ref{fig_parameters}) and also perform a regression to derive a formula that dictates the relationship between the number of qubits and the Hamming weight, with the number of parameters in the Dicke state ansatz.

Taking into account a fixed value for $k$ and varying the number of qubits $n$ in the range $[k+1, d]$, where $d > k+1$, and building the Dicke state ansatz circuit for each value of $n$, we can calculate the number of parameters. Plotting $n_p$ by $n$, we saw a polynomial function of degree 1, then we can write the following expression
\begin{eqnarray}
    n_{p} = mn + b.
\end{eqnarray}
The angular coefficient can be calculated by the following equation
\begin{eqnarray}
    m = \frac{n_p^f-n_p^i}{n_f-n_i},
\end{eqnarray}
computing the value of $m$ with the data obtained from the quantum circuits of Dicke state ansatz, we found that $m = k$, which implies that the angular coefficient is equal to the Hamming weight of the Dicke state. The linear coefficient can be computed using the following expression
\begin{eqnarray}
    b = n_p^i - kn_i,
\end{eqnarray}
and in order to find out if there is a general rule for $b$, we tested different values of $k$ for a fixed $d$ and we found that the value of the linear coefficient is given by
\begin{eqnarray}
    b = \frac{k(k+1)}{2}.
\end{eqnarray}
Thus, combining all the elements computed above, we obtain the following equation for a Dicke state ansatz with $n$ qubits and Hamming weight $k$
\begin{eqnarray}
    n_p = kn - \frac{k(k+1)}{2}.
\end{eqnarray}

\section{Dicke state ansatz on noisy devices}

Once a good ansatz has been identified, it is important to improve its implementation in hardware to improve the results. In reference \cite{wang2024variational}, the authors proposed a Dicke state ansatz version that is more efficient to run in the current noisy quantum hardware, reducing the number of entangling gates and also considering qubits connections constraints. Another way to improve the depth of the Dicke state circuit proposed in \cite{wang2024variational, mukherjee2020preparing}, it is the parallelization of entangling gates, with which we can optimize the circuit depth in order to create quantum circuits which are suitable for the current quantum hardware coherence time.  A topic to highlight is the Dicke state sensitivity to errors of two types: bit-flip and readout errors. For instance, we can measure the impact of bit-flips, in our ansatz, modeling this type of error as described below and manipulating the probability $p$ of occurrence of a bit-flip \cite{nielsen2010quantum}
\begin{eqnarray}
    \rho = (1-p)\rho_0 + p\sigma_x\rho_0\sigma_x,
\end{eqnarray}
where $\rho_0$ represents the density matrix related to the Dicke state ansatz. Then we can use the noisy state $\rho$ to calculate the Hamiltonian expected value through \cite{weinberg2015lectures}
\begin{eqnarray}
    \langle H \rangle = \textrm{tr}(\rho H),
\end{eqnarray}
and with that we will be able to estimate the impact on our objective function. Both errors can destroy the main characteristics of this quantum state, that is, a constant Hamming weight, because depending of the number of bit-flips and in which qubits they will occur, this kind of error can increase or reduce the Hamming weight, which will reflect in adding Hamiltonian terms that will have a great impact on Hamiltonian expectation value. Then, it is expected that the results obtained from current noisy quantum hardware will not achieve the same level of convergence that we achieved with noiseless simulation, because bit-flips and readout errors can make parameter optimization more challenging, since noise can also induce Barren Plateaus \cite{wang2021noise}. 

One last type of error that can cause issues during parameter optimization is the coherent error, more specifically, an error of the form $U(\theta + \delta\theta)$, where $\delta\theta$ represents a displacement in the original parameter. In order to mitigate the damage that the errors described above can cause, we can use some error mitigation techniques. For example, to mitigate errors in expectation values we can use ZNE \cite{kandala2018extending, giurgica2020digital, kim2023scalable}, PEC \cite{van2023probabilistic, gupta2024probabilistic} and Probabilistic Error Amplification (PEA) \cite{filippov2023scalable, kim2023evidence}. For readout errors, we can apply statistical corrections \cite{nation2021scalable, van2022model}, M3 technique \cite{nation2021scalable} or Twirled Readout Error Extinction (T-REX) \cite{van2022model}. The usage of the error mitigation techniques mentioned above, combined with advanced transpiling methods are highly recommended, in order to produce optimized quantum circuits to current hardware, improving the chances of obtaining good results. Another interesting direction in which we would like to explore CVaR as the objective function \cite{barkoutsos2020improving, barron2024provable}, once CVaR seems to be a noise-resistant loss function for variational quantum algorithms.

In order to assess the effectiveness of our simulation results on current quantum hardware, we carried out experiments utilizing the IBM QPU \textit{ibm\_pittsburgh}. These tests were executed in Scenario I, employing the most optimal set of parameters post-optimization. The parameters we identified ensured a $100\%$ probability of sampling the global optimum. We analyze the distribution obtained from the hardware against the simulation, which in this scenario matches the ideal result. As observed in Supplementary Figure~\ref{qpu_histogram}, the hardware results are quite noisy and do not identify the global optima as the bit string with the highest sampling probability. Additionally, it is noteworthy that the Hamming weight of the sampled bit strings failed to maintain the Hamming established by $k=4$, since infeasible solutions appear in the results. This result aligns with the discussion in this section, where it was emphasized that in NISQ hardware, infeasible solutions may arise due to bit-flip and readout errors.
\begin{figure}[h!]
\begin{center}
\includegraphics[scale=.7]{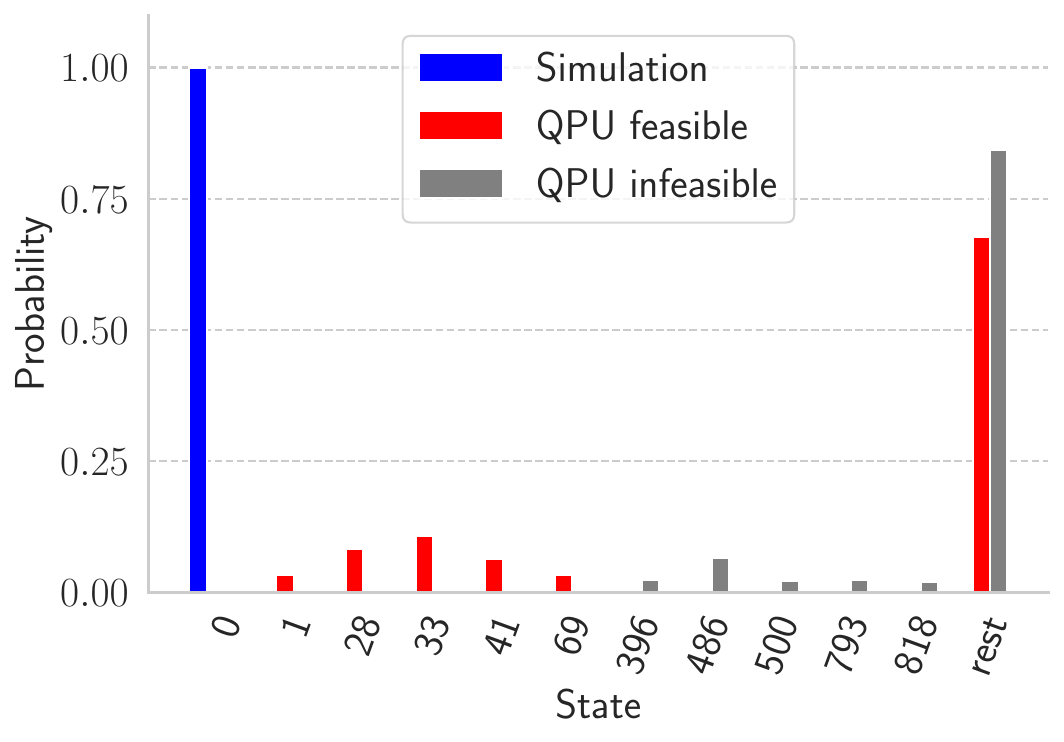}
\end{center}
\caption{Scenario I: A comparison between a simulation using optimal parameters that achieves the ground state with a $100\%$ probability and the distribution obtained from hardware using the identical circuit and parameters. The states are labeled with integer numbers according to their energy levels. Thus, the ground state is labeled as $0$, the first excited state as $1$, and this pattern continues similarly. In blue, we have the simulation results, while the hardware results are in red and gray. The red indicates the feasible bit strings, whereas the gray  the infeasible bit strings, which do not comply with the Hamming weight constraints. Observe that the search space size is $210$, meaning that indexes greater than this indicate infeasible bit strings. The optimization level used in Qiskit Transpiler was 3.}
\label{qpu_histogram}
\end{figure}

Additionally, we calculate the expectation value of the Ising Hamiltonian for Scenario I using the Qiskit Estimator primitive over 100 executions on IBM quantum hardware. We chose the set of parameters that yielded the best outcome in simulations and incorporated these parameters into the Dicke state ansatz. Subsequently, the expectation value was calculated 100 times using the same quantum circuit. The results are shown in Supplementary Figure \ref{energy_distribution}
\begin{figure}[h!]
\begin{center}
\includegraphics[scale=.7]{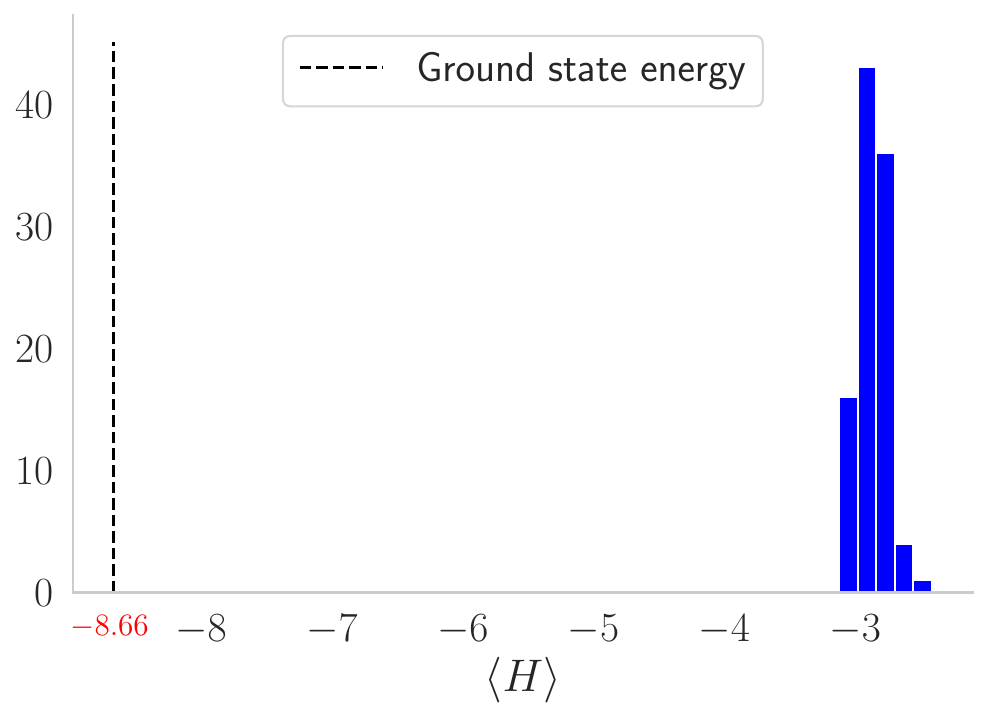}
\end{center}
\caption{Calculation of the expected value of the Hamiltonian using $100$ Qiskit Estimator executions in Scenario I. The vertical dashed black line indicates the ground state energy that we aim to reach. It is evident that the distribution is significantly distant from the optimal regime.}
\label{energy_distribution}
\end{figure}

It is important to note that these experiments were performed using the default settings of Qiskit Sampler and Estimator, and we also considered the standard transpilation techniques adopted in Qiskit transpiler. The results presented above show what we expected to see in noisy hardware without circuit optimizations and error mitigation techniques, however these data do not contradict our simulation results, they reflect the challenge that is extract a result that is comparable with a noiseless simulation. In future research, we would like to address these problems of hardware performance with a more optimized circuit and also using the error mitigation techniques mentioned.

\section{Quantum state fidelity analysis}
\label{appendix_fidelity}

\begin{figure}[h!]
\begin{center}
\includegraphics[scale=.7]{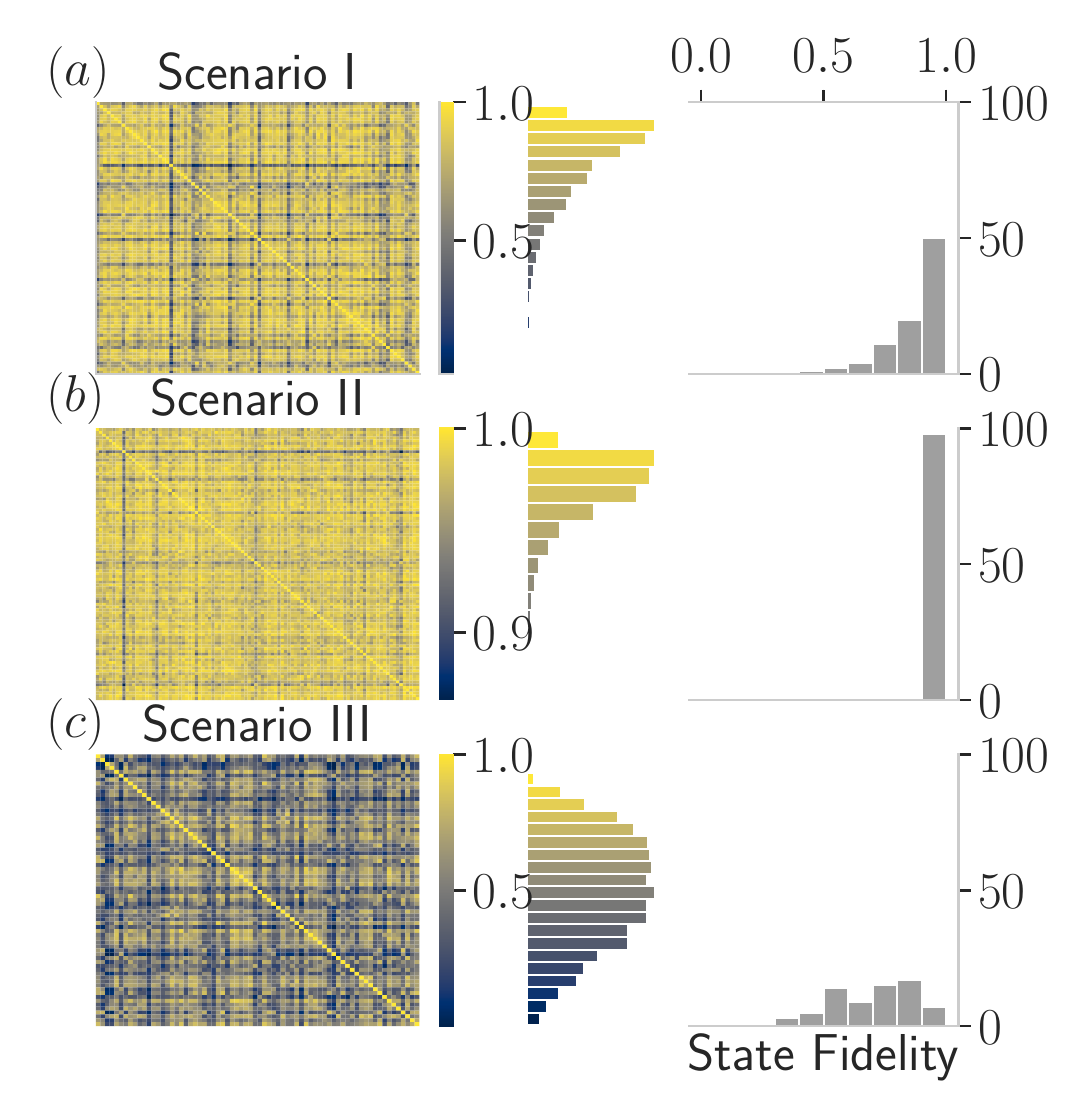}
\end{center}
\caption{A visual analysis of the state fidelity behavior in each scenario proposed. The heatmap represents the state fidelity between of each experiment with Dicke state and CMA-ES, then each point reflects the similarity between the quantum states produced by two sets of parameters obtained at the end of the corresponding experiments. The main diagonal is always equal to 1, because it is the fidelity between the quantum states defined by the same set of parameters. The colored histogram in the Y-axis shows how the points in the heatmap are distributed in the state fidelity scale. The gray histogram represents the state fidelity distribution obtained when we compute the fidelity between the quantum states produced by the optimal set of parameters for each experiments and the reference state for each case: (a) Scenario I results. (b) Scenario II results. (c) Scenario III results.}
\label{fidelity}
\end{figure}

 An additional analysis that we performed was on the quantum state fidelity between states at the end of the optimization process, but also on their state fidelity with reference state. Here we only consider results obtained with Dicke state ansatz simulations using CMA-ES as optimizer, these results are summarized in Figure \ref{fidelity}.

 The state fidelity computation was done considering pure states, then we used the following mathematical expression \cite{nielsen2010quantum} 
\begin{eqnarray}
    F(\vert \psi(\vec{\theta^*_i})\rangle, \vert \psi(\vec{\theta^*_j})\rangle) = \vert\langle \psi(\vec{\theta^*_j})\vert \psi(\vec{\theta^*_i})\rangle\vert^2, \label{fidelity}
\end{eqnarray}
where $\vec{\theta^*_i}$ and $\vec{\theta^*_j}$ represent two different optimal set of parameters for the states described in Table \ref{table:scenarios}. To compute the state fidelity, we consider a parameterized quantum circuit that is used to compute the kernel in Quantum Support Vector Machine (QSVM) \cite{chen2024validating}, where we apply an unitary operator $U(\vec{\theta})$ that prepares the desired state and its adjoint with different set of parameters, as stated in the following expression
\begin{eqnarray}
    \vert\Phi\rangle = U^\dagger(\vec{\theta_j})U(\vec{\theta_i})\vert0\rangle^{\otimes \ n}.
\end{eqnarray}
Then, our estimation of fidelity will be equal to the probability of measuring $\vert0\rangle^{\otimes \ n}$. This fidelity estimation could also be done with the SWAP test \cite{fanizza2020beyond}.  

Our finding here was that these different sets of optimal parameters lead to quantum states that have a high fidelity to the reference state. We also noted a certain level of similarity between these quantum states, as shown by the heat maps plots in the Supplementary Figure \ref{fidelity}. However, we cannot extract useful information from the distance between the parameter sets, because it is not clear how we can define what is close or distant in the parameter space. 

Perhaps, with this information in hand, we can extract some useful information about the optimization landscape to improve the success of this method. A future research line would design a neural network to provide a good initial starting set of parameters considering this information about the similarity between the states that have a high fidelity with the reference state and are similar between them. 


\end{document}